\documentclass[structabstract]{aa}  
                                                                    
\usepackage{graphicx}
\usepackage{natbib} 
\usepackage[utf8]{inputenc}

\usepackage{txfonts}
\usepackage{bm}
\usepackage{url}
\usepackage{xcolor}
\usepackage{multirow}
\usepackage[normalem]{ulem}
\newcommand{\emm}[1]{\ensuremath{#1}}
\newcommand{\emr}[1]{\emm{\mathrm{#1}}}
\newcommand{\unit}[1]{\emr{\,#1}}
\newcommand{\pc}{\unit{pc}}

\newcommand{\pccm}{\unit{cm^{-3}}}
\newcommand{\ccm}{\unit{cm^{3}}}
\newcommand{\pscm}{\unit{cm^{-2}}}
\newcommand{\ps}{\unit{s^{-1}}}
\newcommand{\kms}{\unit{km\,s^{-1}}}
\newcommand{\K}{\unit{K}}
\newcommand{\mK}{\unit{mK}}
\newcommand{\mm}{\unit{mm}}
\newcommand{\kHz}{\unit{kHz}}
\newcommand{\GHz}{\unit{GHz}}

\newcommand{\chem}[1]{\ensuremath{\mathrm{#1}}}

\newcommand{\metcy}{\chem{CH_{3}CN}}
\newcommand{\isometcy}{\chem{CH_{3}NC}}
\newcommand{\cyacet}{\chem{HC_{3}N}}
\newcommand{\isocyacet}{\chem{HC_{2}NC}}
\newcommand{\hnccc}{\chem{HNC_{3}}}
\newcommand{\cycethy}{\chem{C_{3}N}}

\newcommand{\Ht}{\chem{H_{2}}}
\newcommand{\DCOp}{\chem{DCO^{+}}}

\newcommand{\Tkin}{\emm{T_\emr{kin}}}

\newcommand{\nH}{\emm{n_\emr{H}}}

\newcommand{\Tmb}{\emr{T_{mb}}}
\newcommand{\Vlsr}{\unit{V_\emr{LSR}}}
\newcommand{\dV}{\emm{\Delta V}}
\newcommand{\Eu}{\unit{E_u}} 
\newcommand{\gu}{\unit{g_u}} 
\newcommand{\Au}{\unit{A_u}} 

\newcommand{\Jq}{\emm{J}} 
\newcommand{\Kq}{\emm{K}}
\newcommand{\Ae}{\emm{A_\emr{e}}}
\newcommand{\Be}{\emm{B_\emr{e}}}
\newcommand{\D}[1]{\emm{D_{#1}}} 
\newcommand{\EJK}{\emm{E_{\Jq,\Kq}}}
\newcommand{\JK}[3]{\emm{J=#1\rightarrow#2,\,K=#3}}
\newcommand{\JKsimple}[3]{\emm{#1\rightarrow#2,\,#3\rightarrow#3}}
\newcommand{\J}[2]{\emm{J=#1\rightarrow#2}}

\newcommand{\Jo}[2]{\emm{#1\rightarrow#2}}
\newcommand{\Ko}[2]{\emm{#1\rightarrow#2}}

\newcommand{\sciexp}[2]{\emm{#1\times10^{#2}}}
\newcommand{\radec}[6]{RA=$#1^{h}#2^{m}#3^{s}$, DEC=$#4^{\circ}#5^{'}#6^{''}$}

\newcommand{\ie} {{\em i.e.}}
\newcommand{\eg} {{\em e.g.}}
\newcommand{\etal} {et al.}
\newcommand{\dix}[1]{\times 10^{#1}}

\newcommand{\FigHorseheadOverview}{  \begin{figure*}
    \begin{center}
      \includegraphics[width=\hsize{}]{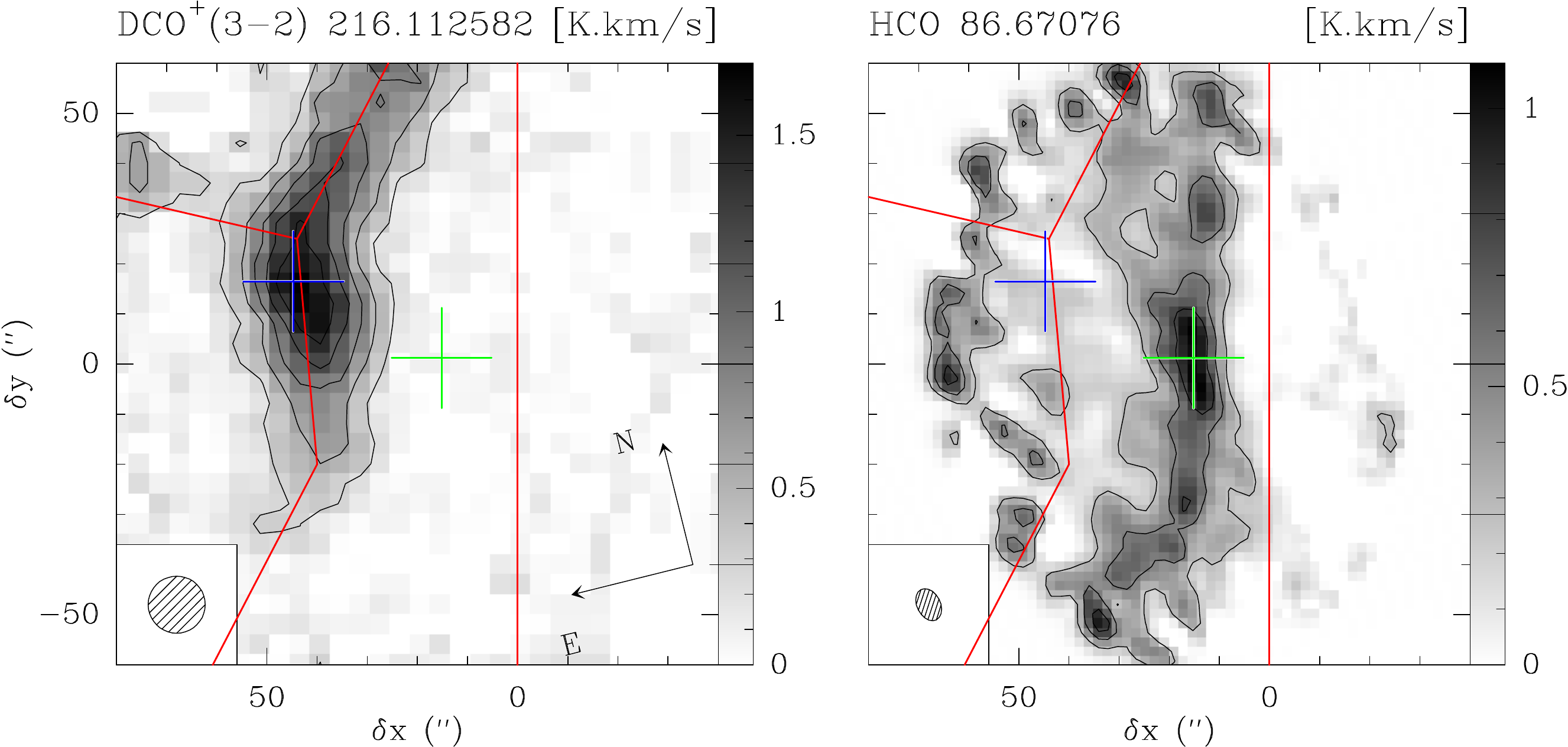}
    \end{center}
    \caption{\label{fig.horsehead_overview} Overview of the Horsehead mane 
      photodissociation region showing the two observed positions.
      \emph{Left:} \DCOp{} map \citep{Pety.2007}, \emph{right:} HCO
      emission map \citep{Gerin.2009a}. In both maps, The HCO emission peak
      (PDR position) at \radec{05}{40}{53.9}{-02}{28}{00} is shown with a
      green cross, and the \DCOp{} emission peak (dense core position) at
      \radec{05}{40}{55.7}{-02}{28}{22} with a blue one. The maps have been
      rotated 14 degree anti clockwise around the position
      \radec{05}{40}{54.27}{-02}{28}{00} and shifted $20''$ to the east to align the
      PDR front (vertical red line) with the horizontal 0 offset of the
      map.}
  \end{figure*}}

\newcommand{\FigMetcyEdiagram}{  \begin{figure}
    \begin{center}
      \includegraphics[width=88mm]{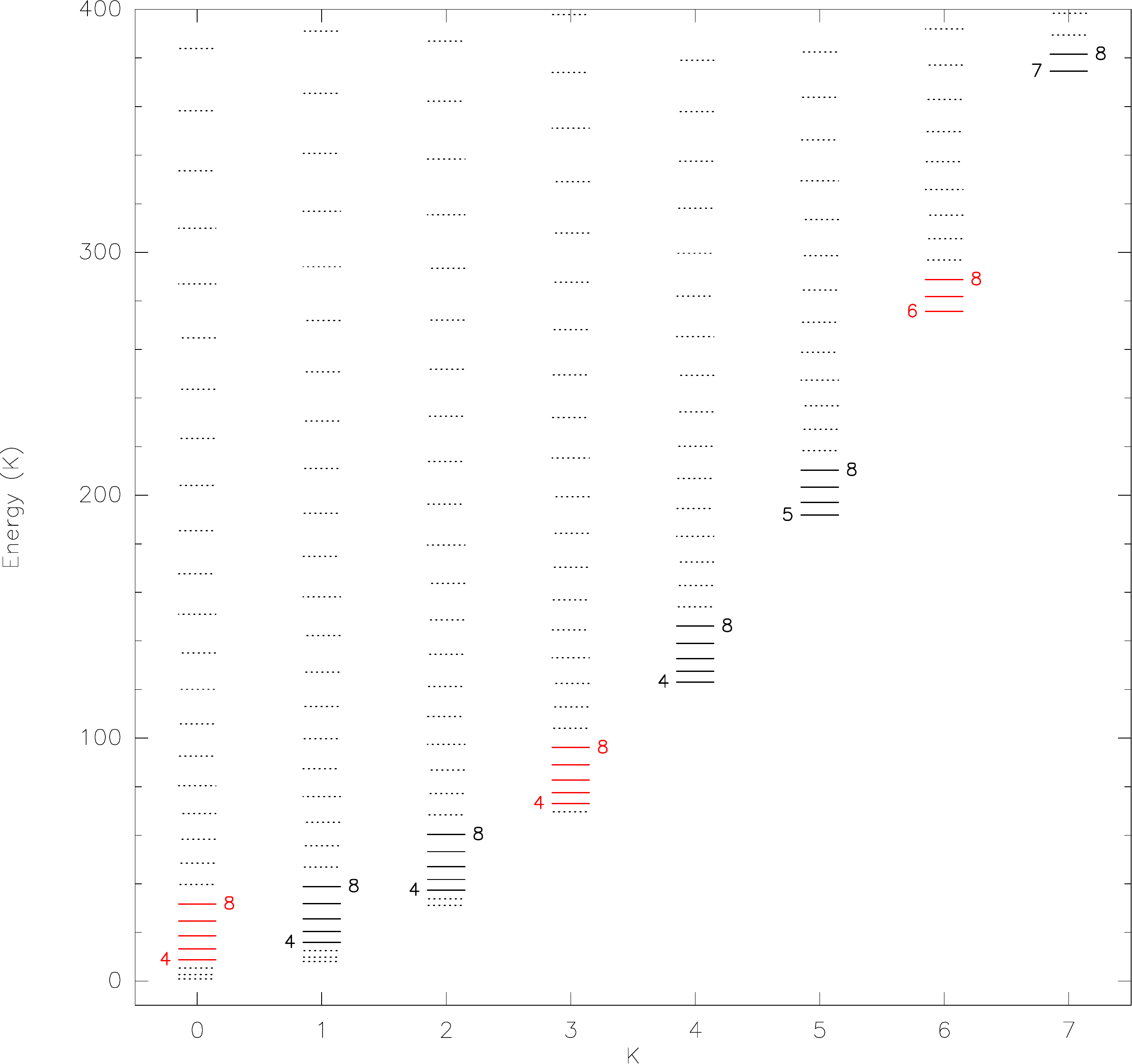}
    \end{center}
    \caption{\metcy{} energy diagram of levels which contribute to the $\Jq{}=\Jo{5}{4}$, $\Jq{}=\Jo{6}{5}$, $\Jq{}=\Jo{7}{6}$, and $\Jq{}=\Jo{8}{7}$ transitions in
      the 3 and 2mm frequency ranges, \ie{} for $K$ levels from 0 to 7 and
      $J$ levels from $4$ to 8, other levels are displayed in { dotted line}. The E
      ($K= 3n\pm1$) and A ($K = 3n$) states are displayed in black and red,
      respectively.}
    \label{fig.ch3cn_energy}
  \end{figure}}

\newcommand{\FigIsometcySpectra}{  \begin{figure*}
    \begin{center}
      \includegraphics[width=\textwidth]{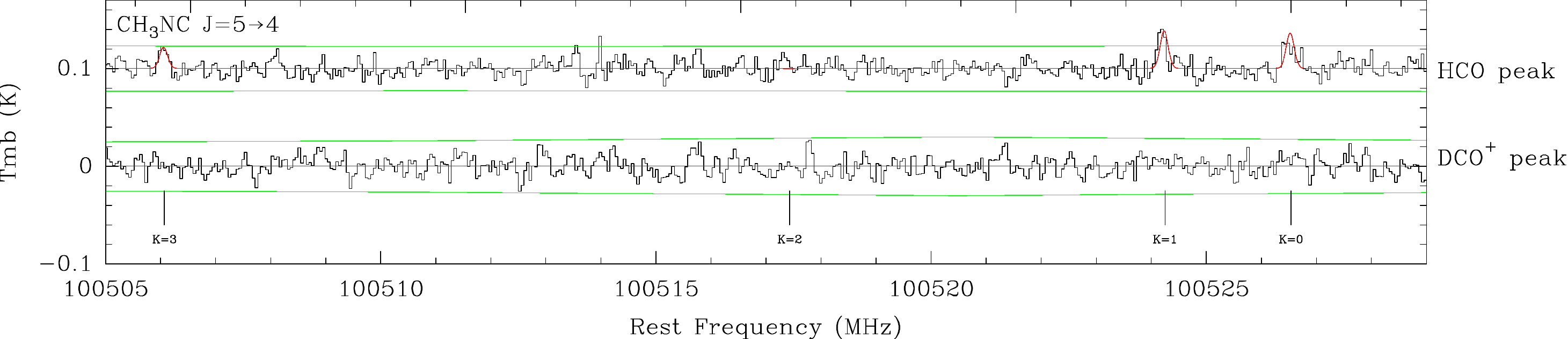}
      \caption{Spectrum of the \isometcy{} 3mm detected lines at the HCO peak
        (\ie{}, the PDR position, top spectrum) and the \DCOp{} peak
        (\ie{}, inside the cold dense core, bottom spectrum), the spectrum towards the HCO peak has been shifted vertically by 0.1~K
for clarity. The panel
        displays the $K$ set of lines for the given $\Delta J$ set of
        lines.  The frequencies corresponding to each transitions are
        displayed as vertical bars. The best fit model is overplotted in
        red. The green horizontal lines display the $\pm3\sigma$
        significance levels.}
      \label{fig.ch3nc_spec}
    \end{center}
  \end{figure*}} 
  
\newcommand{\FigIsometcySpectraSideway}{  \begin{figure*}
    \begin{center}
      \includegraphics[width=\textheight,angle=90]{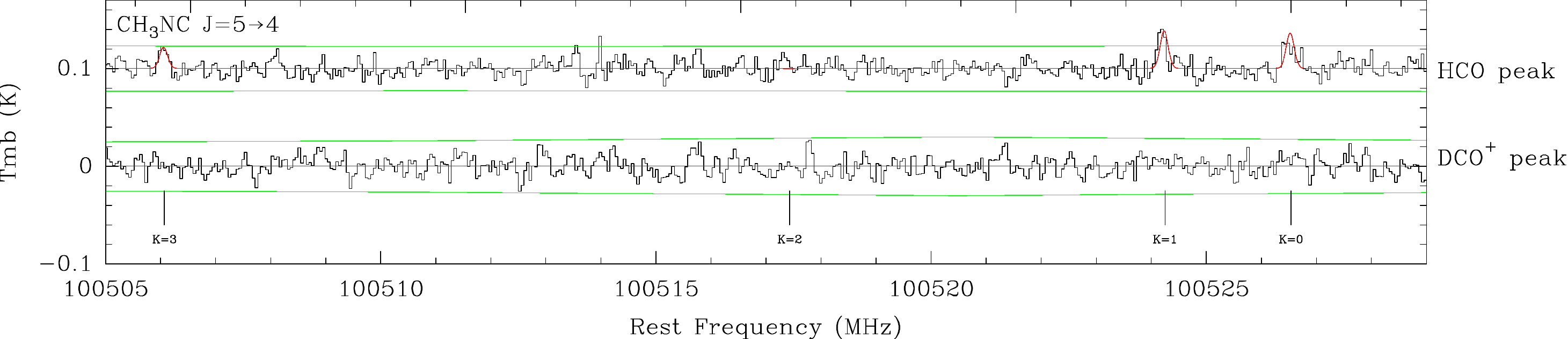}
      \caption{Spectrum of the \isometcy{} 3mm detected lines at the HCO peak
        (\ie{}, the PDR position, top spectrum) and the \DCOp{} peak
        (\ie{}, inside the cold dense core, bottom spectrum), the spectrum towards the HCO peak has been shifted vertically by 0.1~K
for clarity. The panel
        displays the $K$ set of lines for the given $\Delta J$ set of
        lines.  The frequencies corresponding to each transitions are
        displayed as vertical bars. The best fit model is overplotted in
        red. The green horizontal lines display the $\pm3\sigma$
        significance levels.}
      \label{fig.ch3nc_spec_large}
    \end{center}
  \end{figure*}}

\newcommand{\FigMetcySpectra}{  \begin{figure*}
    \begin{center}
      \includegraphics[width=\textwidth]{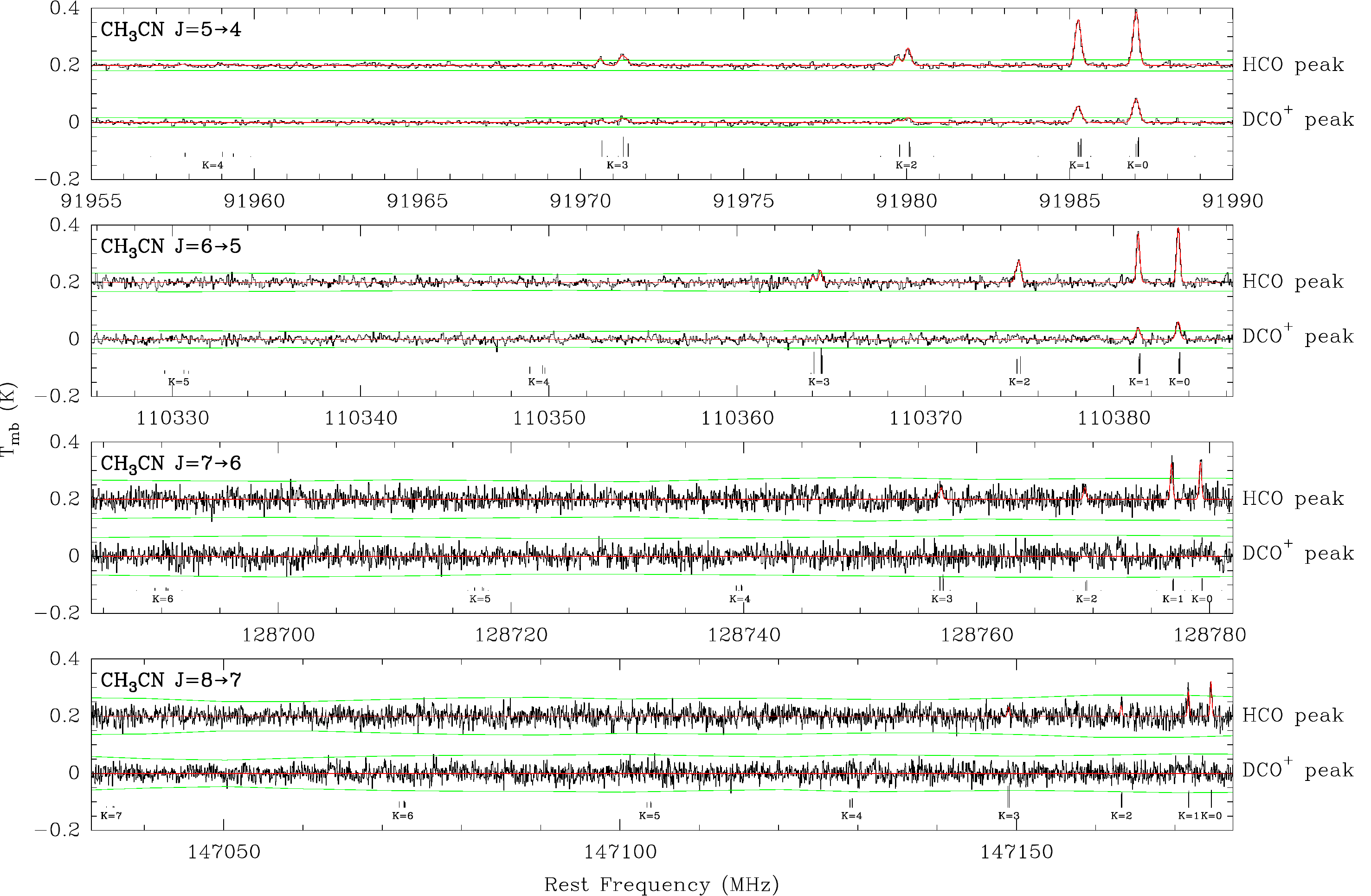}
      \caption{Spectrum of the \metcy{} 3\mm{} and 2\mm{} detected lines at the HCO peak
        (\ie{}, the PDR position, top spectrum of each panel) and the
        \DCOp{} peak (\ie{}, inside the cold dense core, bottom spectrum of
        each panel), each spectrum towards the HCO peak has been shifted vertically by 0.2~K
for clarity. The panel displays the $K$ set of lines for a given
        $\Delta J$ set of lines.  The frequencies corresponding to each
        transitions are displayed as vertical bars, whose heights indicate
        their relative hyperfine intensities in the optically thin regime
        applicable to these observations. The best fit model is overplotted
        in red. The green horizontal lines display the $\pm3\sigma$
        significance levels.}
      \label{fig.ch3cn_spec}
    \end{center}
  \end{figure*}}
  
\newcommand{\FigMetcySpectraSideway}{  \begin{figure*}
    \begin{center}
      \includegraphics[width=\textheight,angle=90]{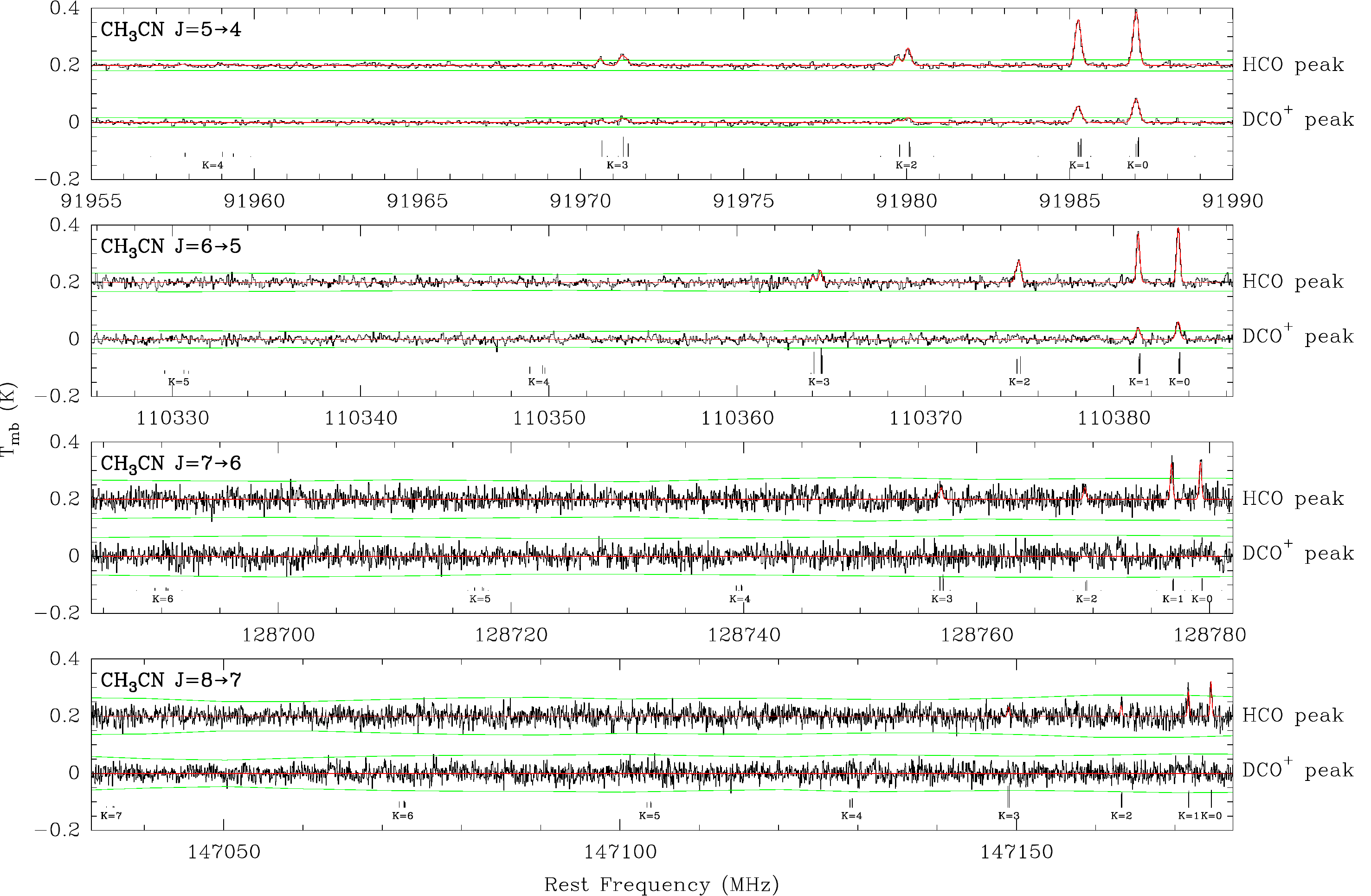}
      \caption{Spectrum of the \metcy{} 3\mm{} and 2\mm{} detected lines at the HCO peak
        (\ie{}, the PDR position, top spectrum of each panel) and the
        \DCOp{} peak (\ie{}, inside the cold dense core, bottom spectrum of
        each panel), each spectrum towards the HCO peak has been shifted vertically by 0.2~K
for clarity. The panel displays the $K$ set of lines for a given
        $\Delta J$ set of lines.  The frequencies corresponding to each
        transitions are displayed as vertical bars, whose heights indicate
        their relative hyperfine intensities in the optically thin regime
        applicable to these observations. The best fit model is overplotted
        in red. The green horizontal lines display the $\pm3\sigma$
        significance levels.}
      \label{fig.ch3cn_spec_large}
    \end{center}
  \end{figure*}}

\newcommand{\FigCyanoacetSpectra}{  \begin{figure*}
    \begin{center}
      \includegraphics[width=\hsize]{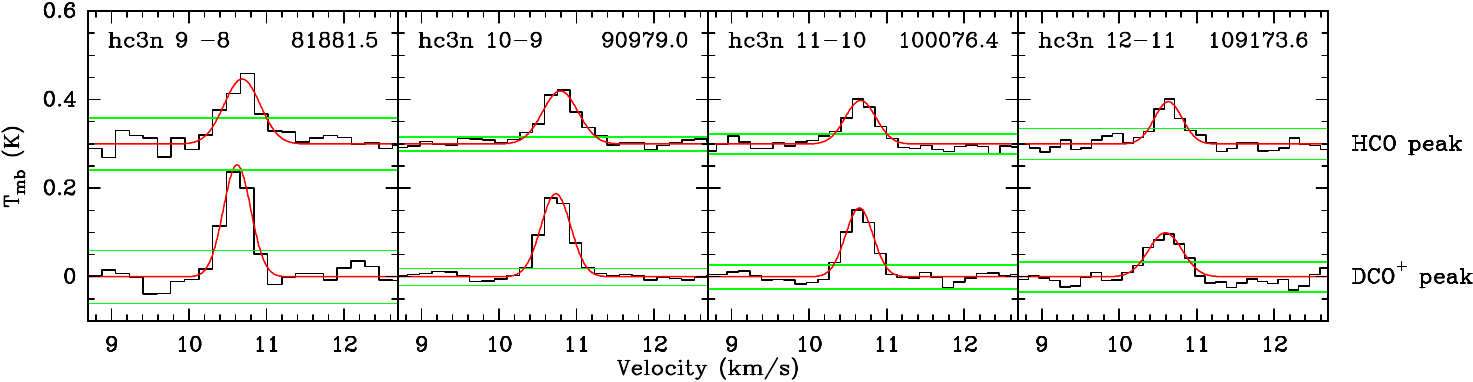}
      \caption{Spectrum of the \cyacet{} 3\mm{}  lines at the HCO peak
        (\ie{}, the PDR position, top spectrum of each panel) and the
        \DCOp{} peak (\ie{}, inside the cold dense core, bottom spectrum of
        each panel), each spectrum towards the HCO peak has been shifted vertically by 0.3~K
for clarity.}
      \label{fig.hc3n_spec}
    \end{center}
  \end{figure*}}
  
\newcommand{\FigCCCNSpectraLTE}{  \begin{figure*}
    \begin{center}
      \includegraphics[width=\hsize]{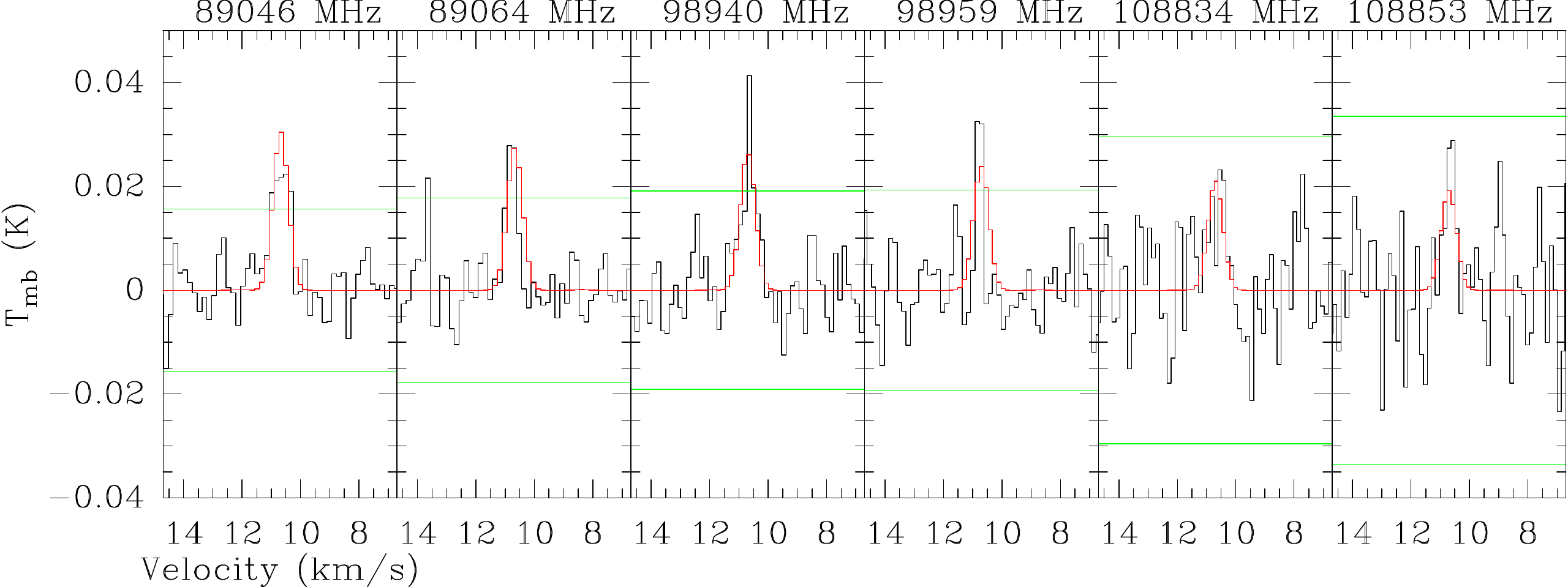}
      \caption{\label{fig.c3n_lte}Observed spectra (\emph{black line}) 
        of the three 3mm \cycethy{} doublets at the PDR peak. The green
        horizontal lines are the $\pm 3\sigma$ levels. The red spectrum is
        the LTE model for a $6'' \times 50''$ filament centered at the PDR
        positions, a 10K excitation temperature, and a column density of
        \sciexp{2}{12}\pscm{}.}
    \end{center}
  \end{figure*}}

\newcommand{\FigPDRDDParameters}{  \begin{figure}
    \begin{center}
      \includegraphics[width=88mm]{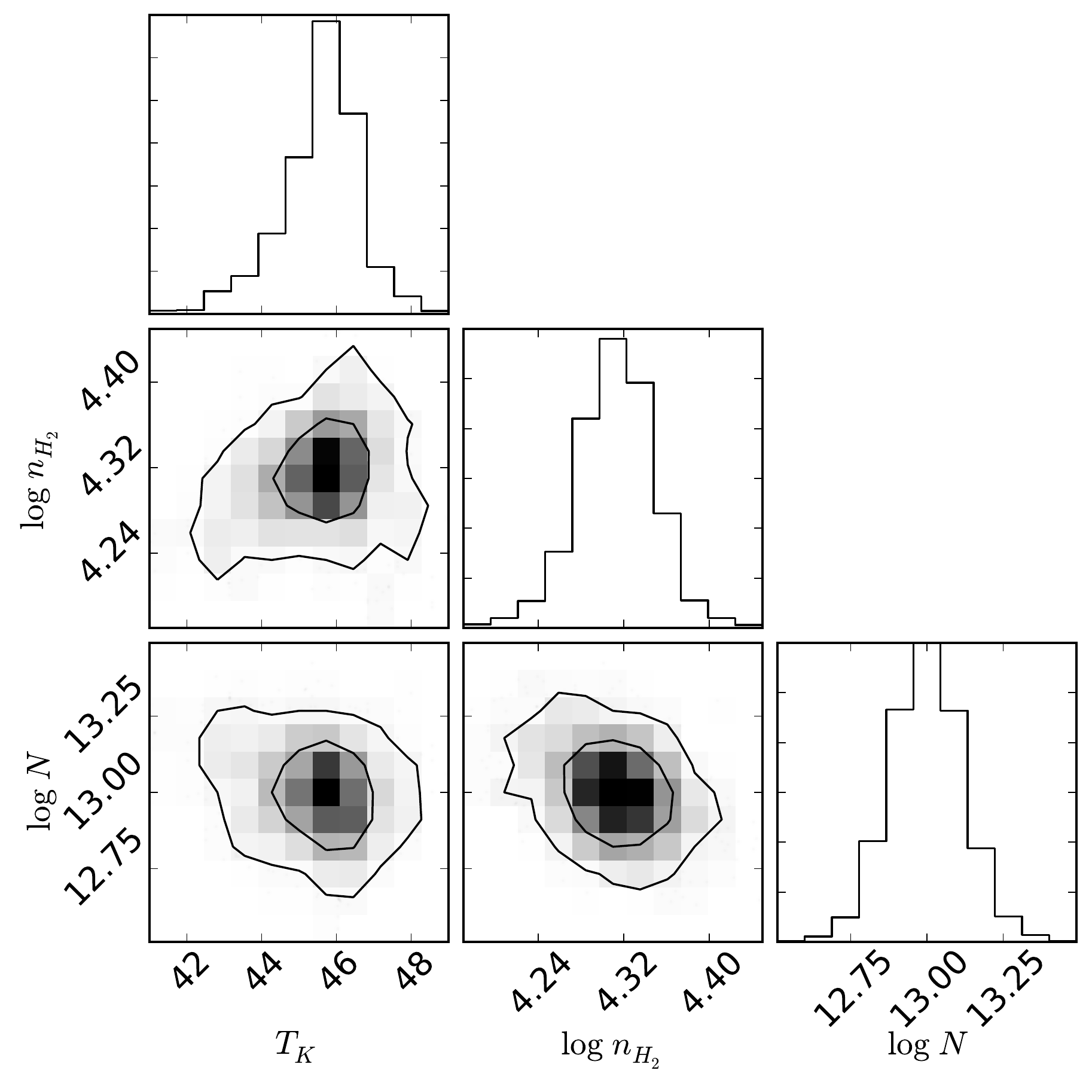}
      \caption{\label{ch3cn_pdr_2dparameters} Distributions of the
        posterior probability for three parameters, \ie{}, gas density
        ($n_{\Ht}$), kinetic temperature ($T_{K}$) and \metcy{} column
        density ($N$), at the PDR position. Along the diagonal, the one
        dimensional probability distribution functions are integrations of
        the two dimension probability distribution functions displayed
        below.  The color coding of the two dimensional histograms runs
        from 0\% (white) to 100\% of the peak value(black). The contours
        correspond to 68\% ($1\sigma$) and 95\% ($2\sigma$) of cumulated
        posterior probability.}
    \end{center}
  \end{figure}}

 \newcommand{\FigPDRIntensitiesTau}{  \begin{figure*}
    \begin{center}
      \includegraphics[width=0.8\textwidth]{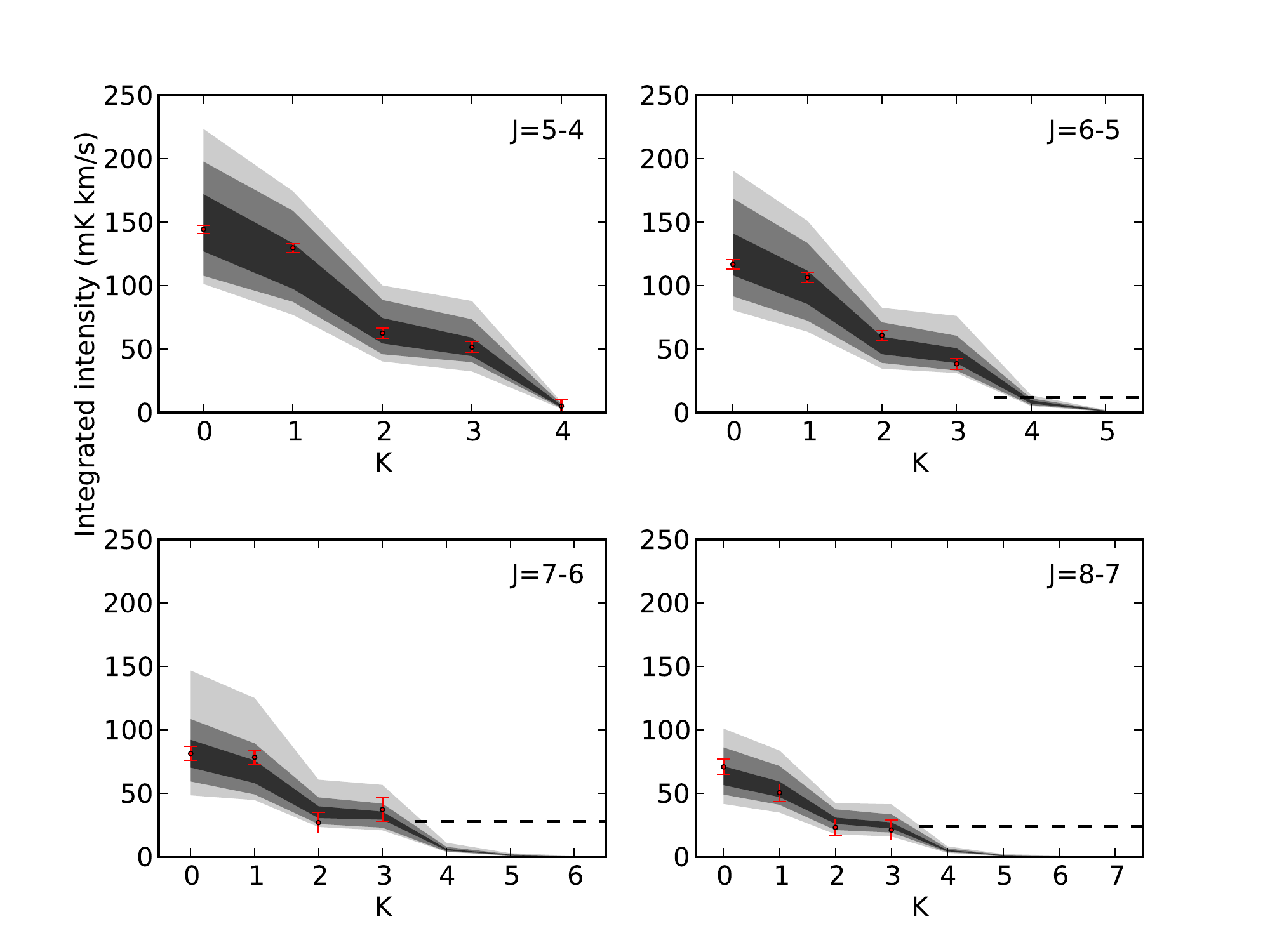}
      \includegraphics[width=0.8\textwidth]{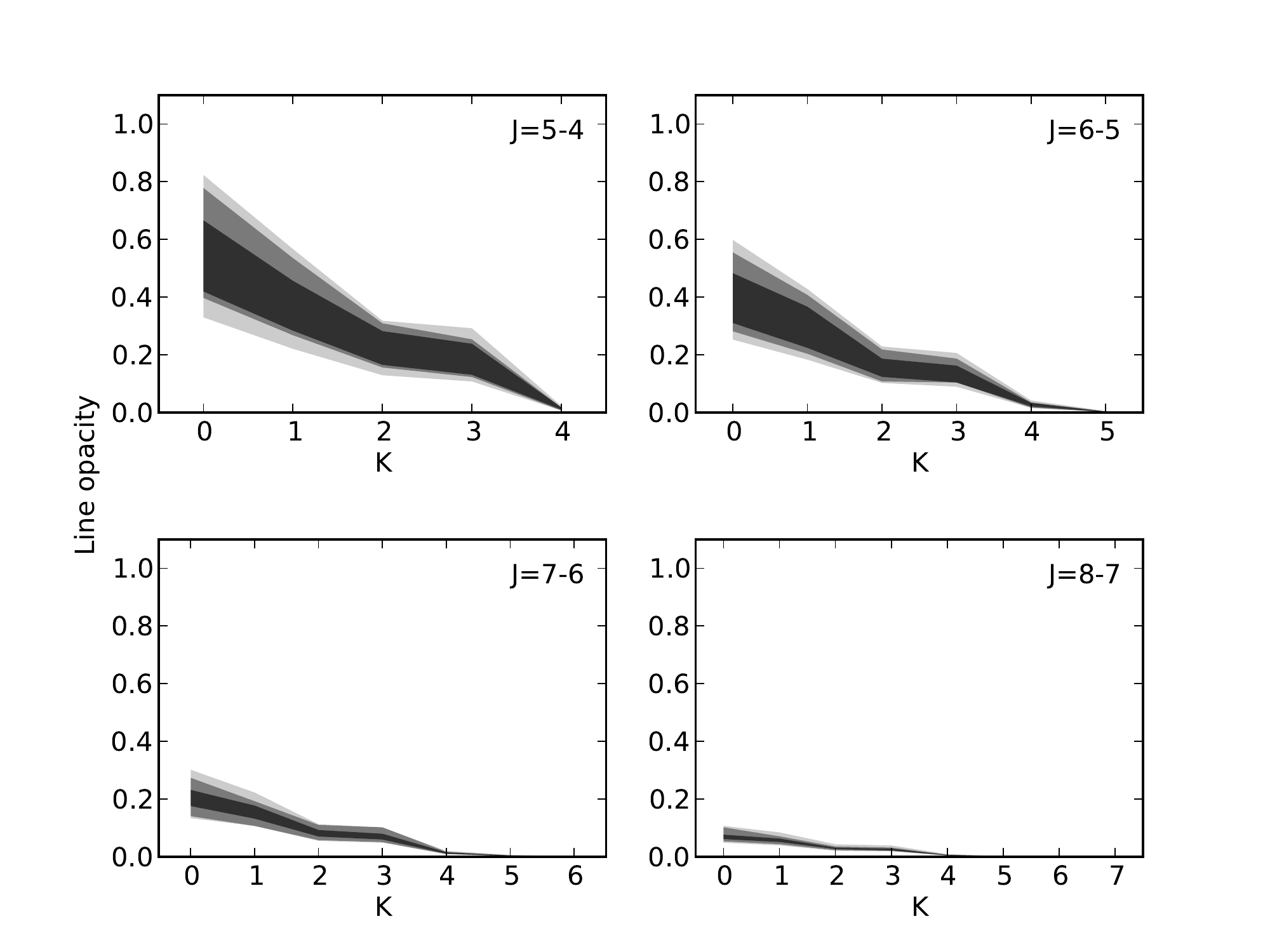}
      \caption{\label{ch3cn_pdr_result}  Distribution
        of modeled integrated intensities (\emph{Top 4 panels}) and line
        opacities (\emph{Bottom 4 panels}) as a function of the K{} number.
        Each panel presents the results for a different (\Jo{J+1}{J})
        \Kq{}-ladder. The 3 different gray levels corresponds to 3
        different uncertainty intervals, \ie{}, 68\% ($1\sigma$), 95\%
        ($2\sigma$), and 99.9\% ($3\sigma$) from dark to light gray.  For
        the 4 top panels, the observed line intensities with their
        $1\sigma$ uncertainty intervals are plotted as red segments The
        dashed horizontal black line displays the $2\sigma$ upper limits for
        the undetected lines.}
    \end{center}
  \end{figure*}}

\newcommand{\FigRotDiagPDR}{  \begin{figure*}
    \begin{center}
      \includegraphics[width=.48\hsize]{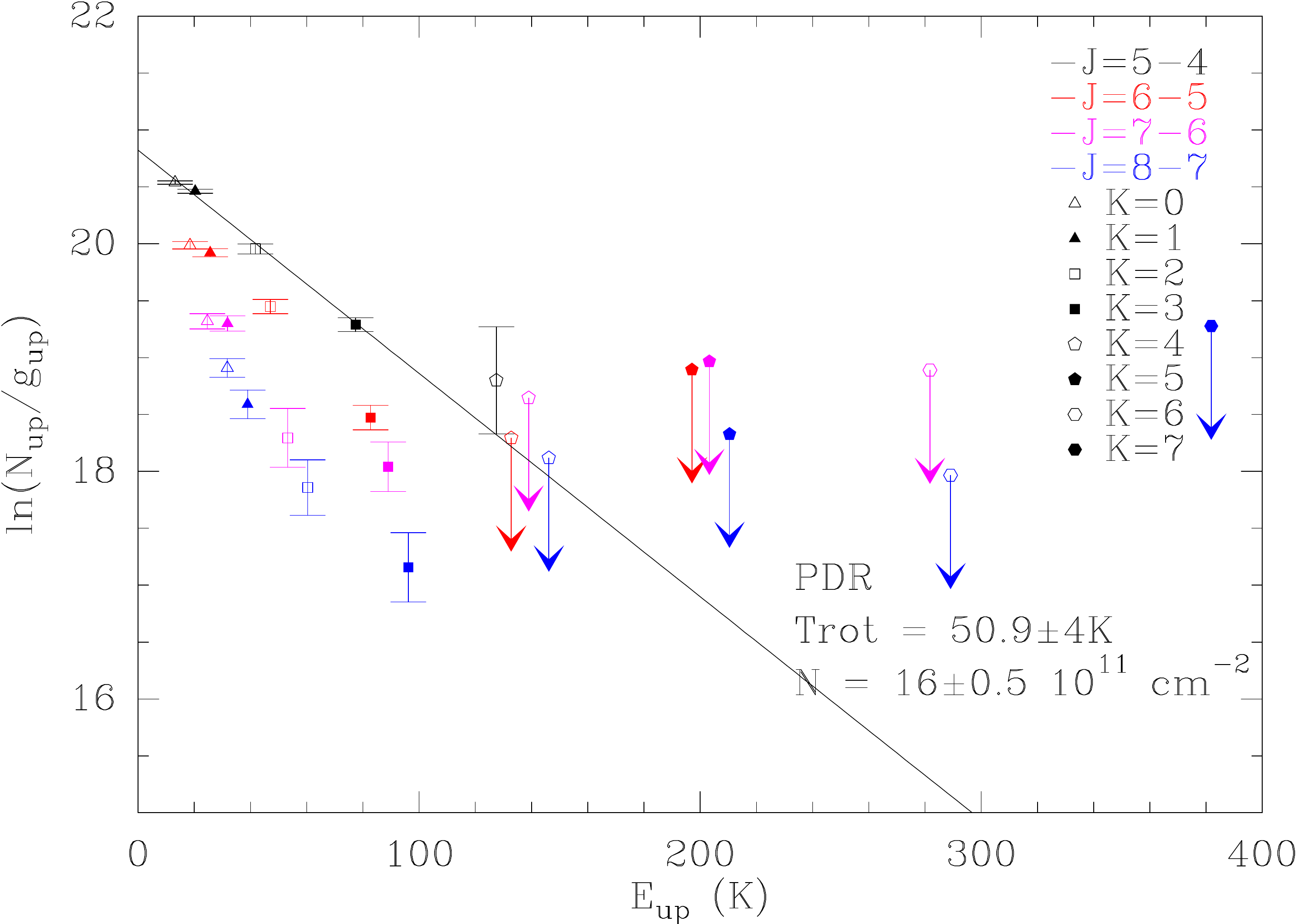}
      \includegraphics[width=.48\hsize]{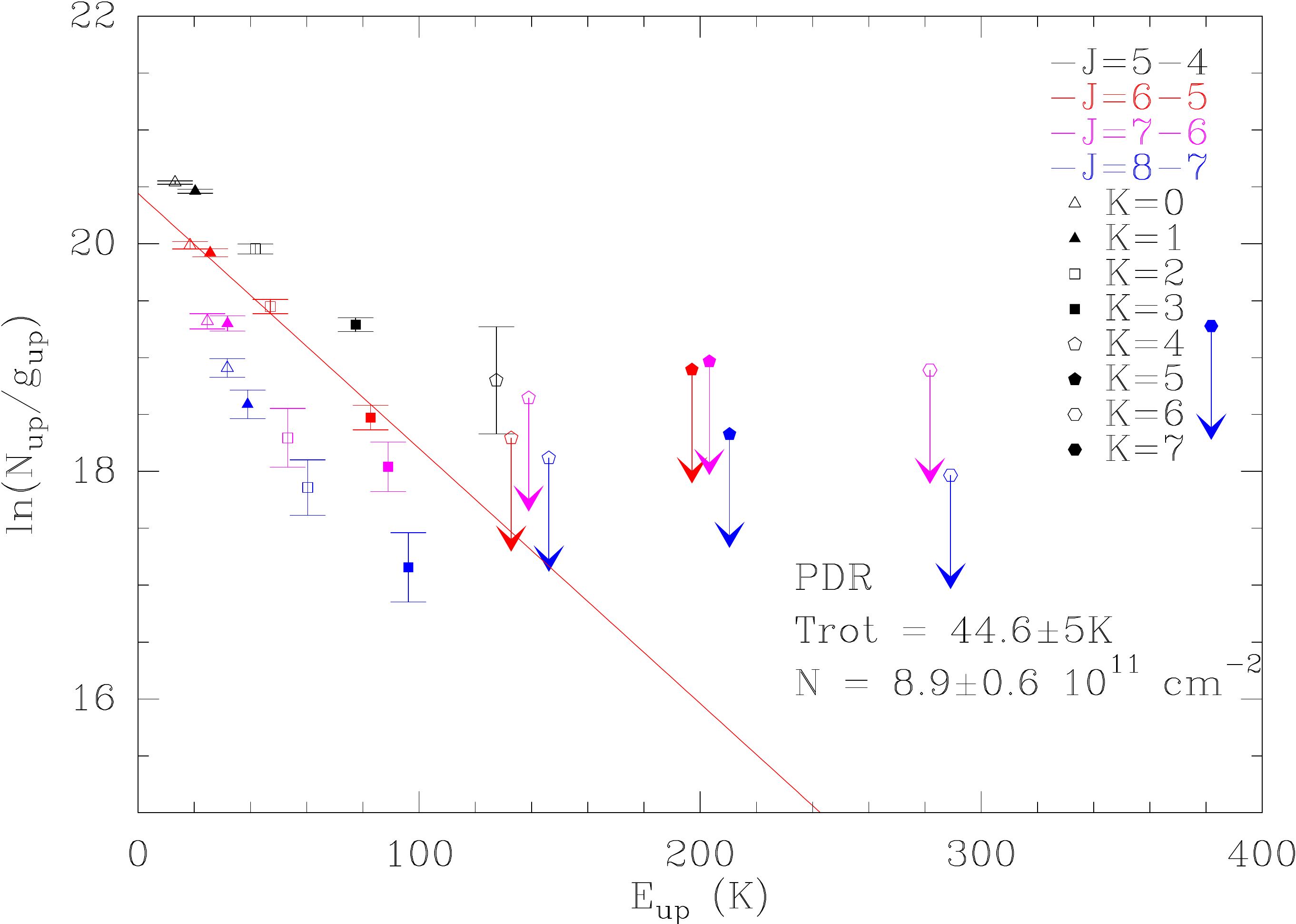}
      \includegraphics[width=.48\hsize]{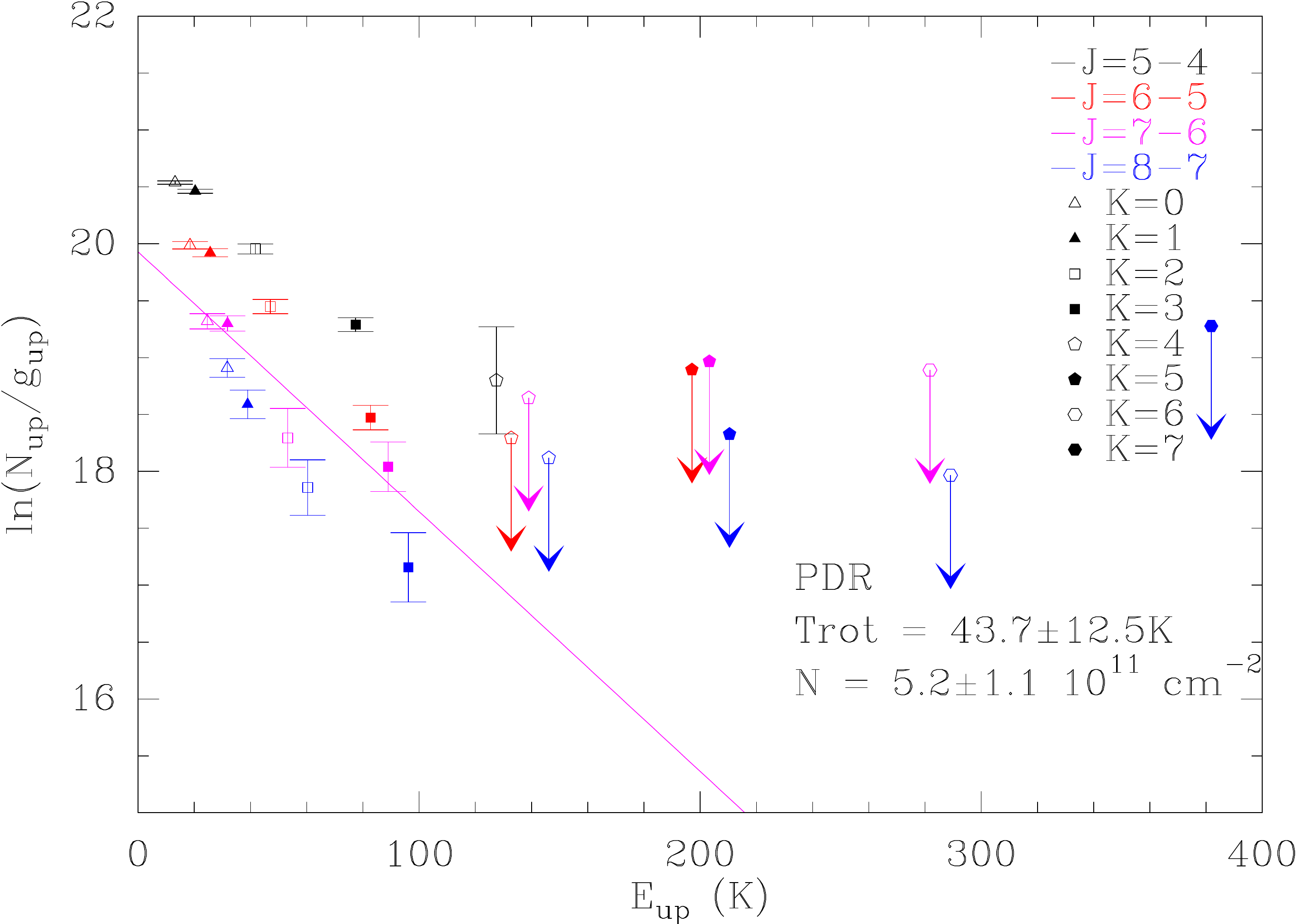}
      \includegraphics[width=.48\hsize]{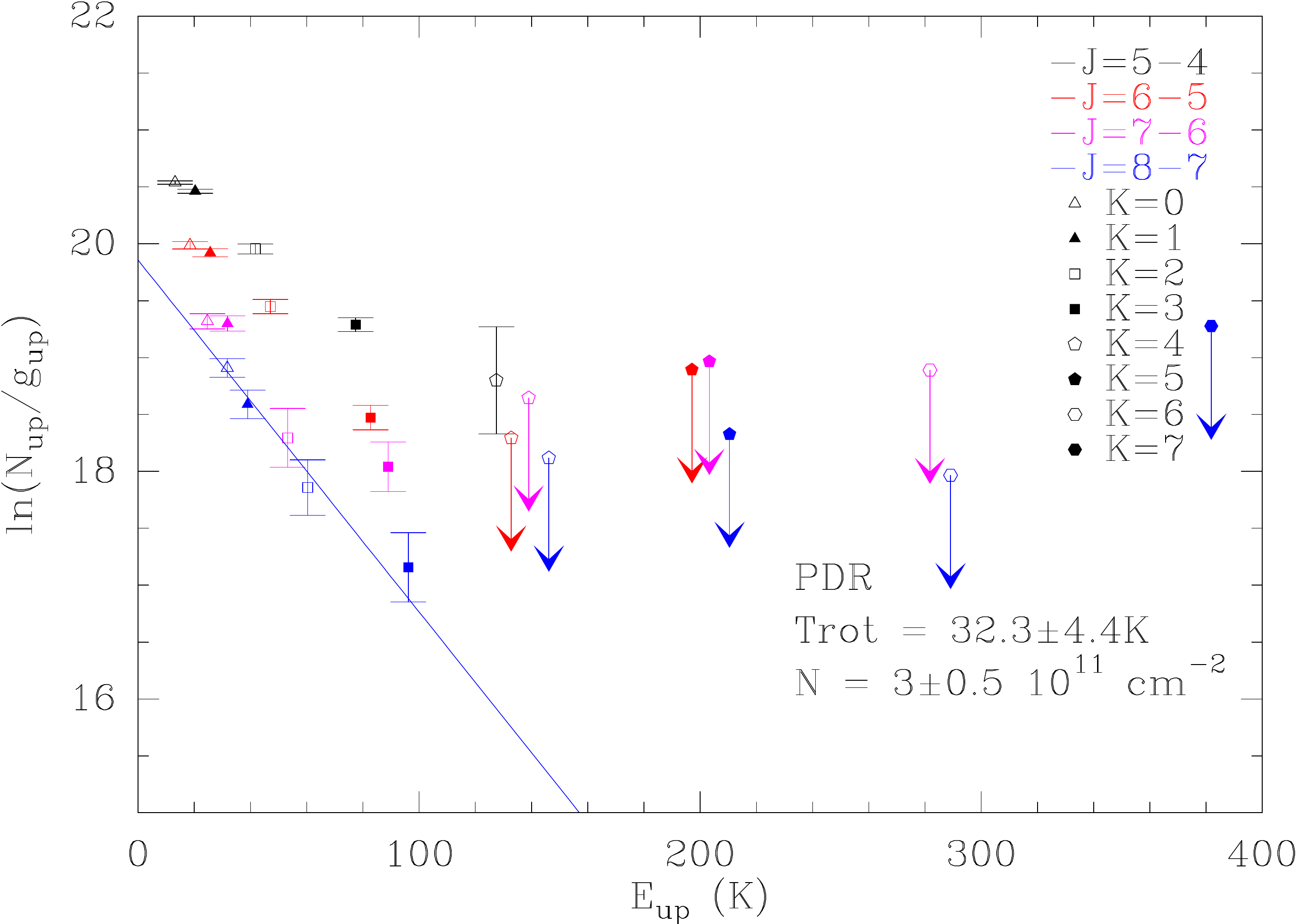}
      \includegraphics[width=.48\hsize]{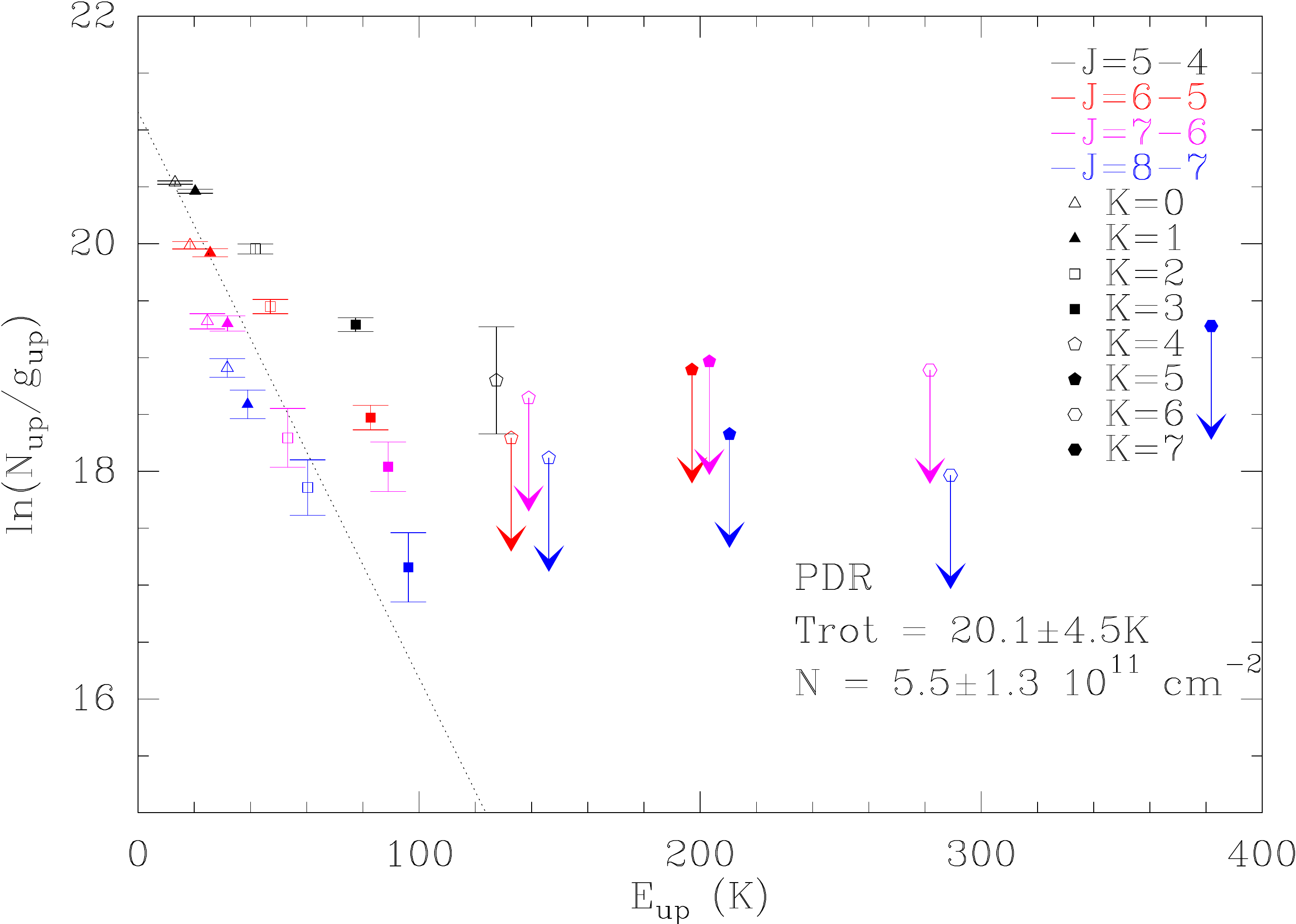}
      \caption{\label{ch3cn_rotdiag_pdr} { Rotational diagrams for \metcy{} at the HCO peak, in each of the 4 top panels, the solid lines  correspond to fits restricted to 
        to  observed \Kq{} transitions for a given \Jq{}--\Jq{}-1 set,
        the fitted line is of the same color as the fitted points.  
        The final panel shows, in a dotted line, the result of fitting all lines simultaneously. Each
        panel shows the  rotational temperature and column density derived from the fit.}}
    \end{center}
  \end{figure*}}
  
\newcommand{\FigRotDiagCore}{  \begin{figure*}
    \begin{center}
      \includegraphics[width=.48\hsize]{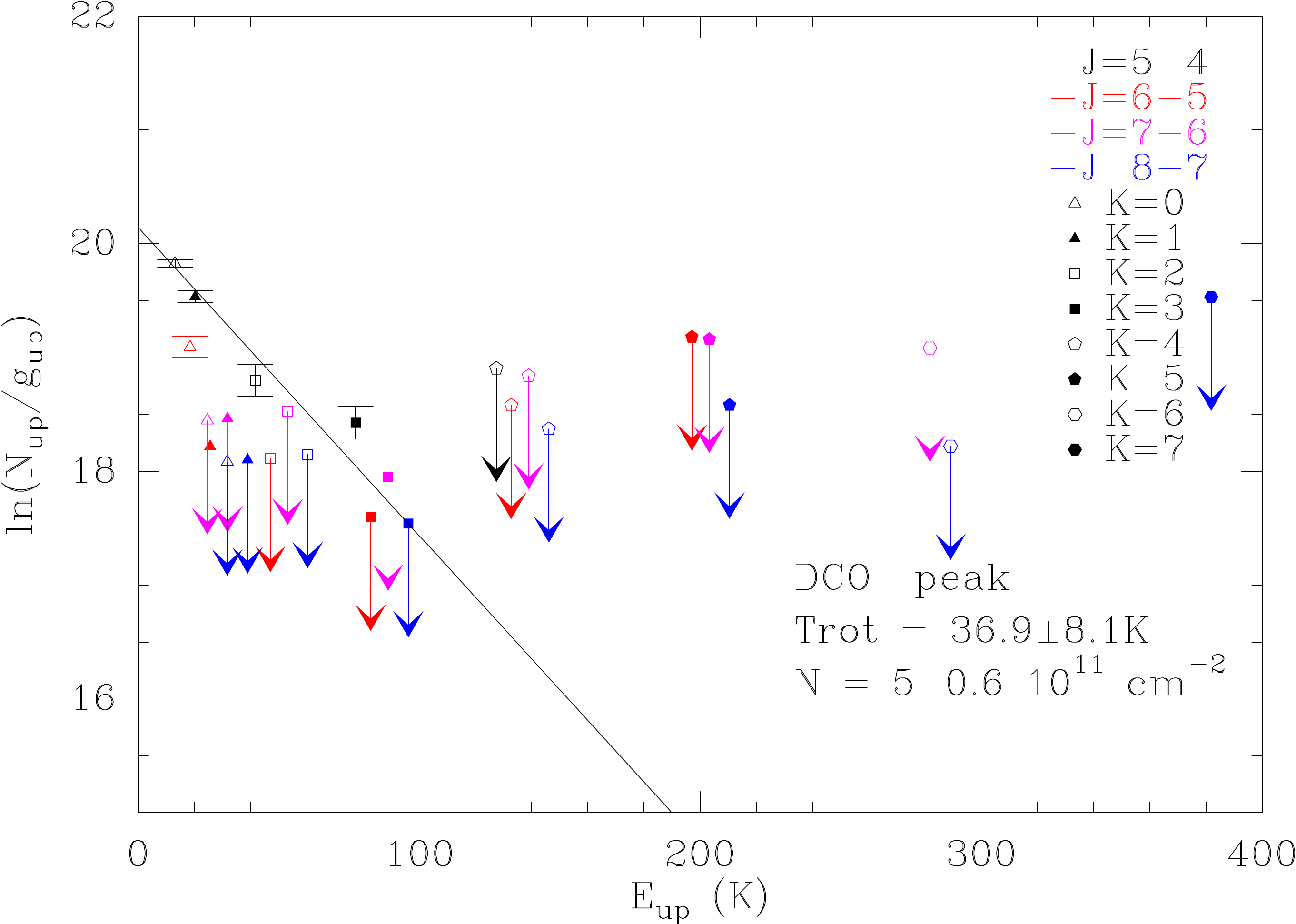}
      \includegraphics[width=.48\hsize]{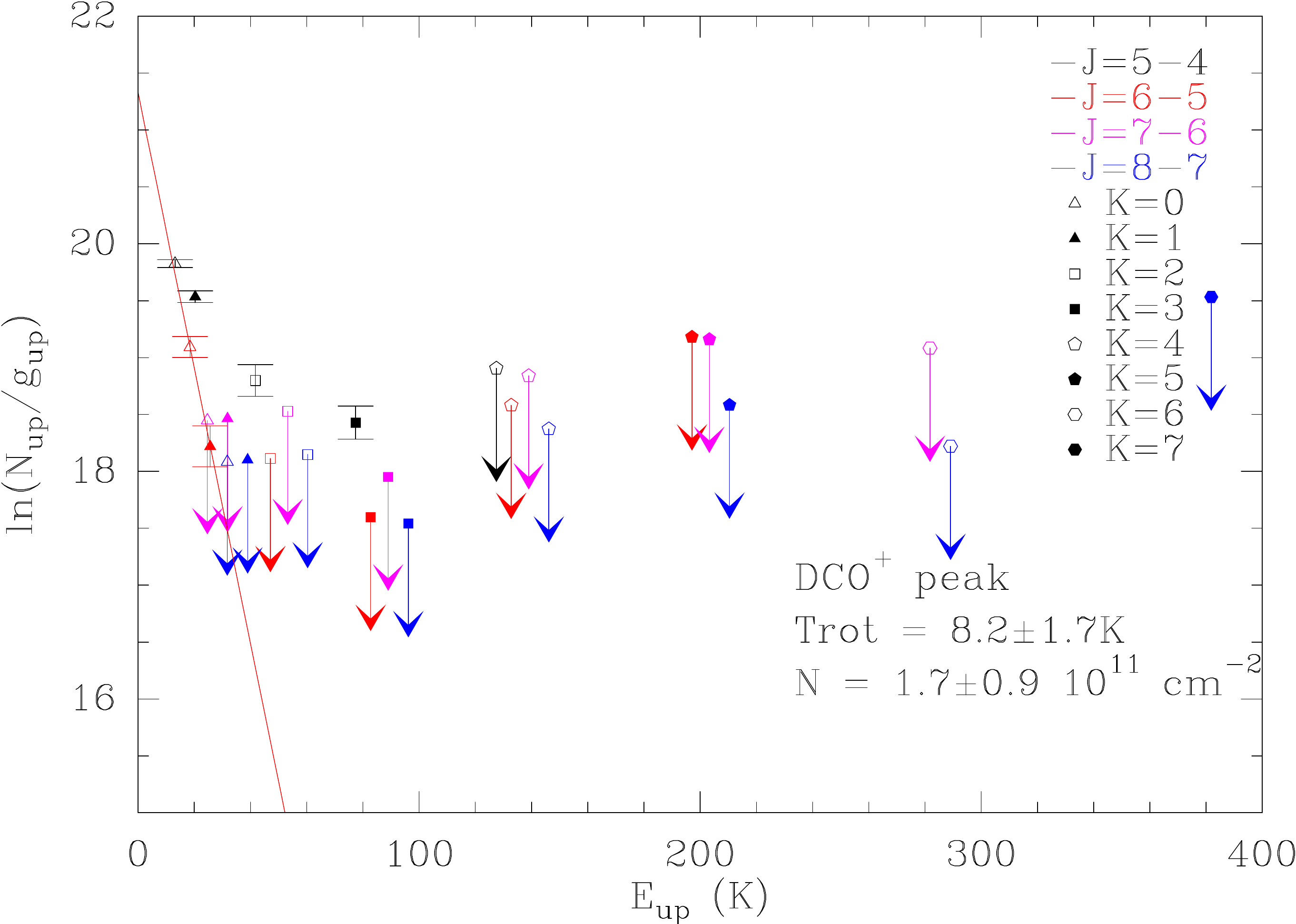}
      \includegraphics[width=.48\hsize]{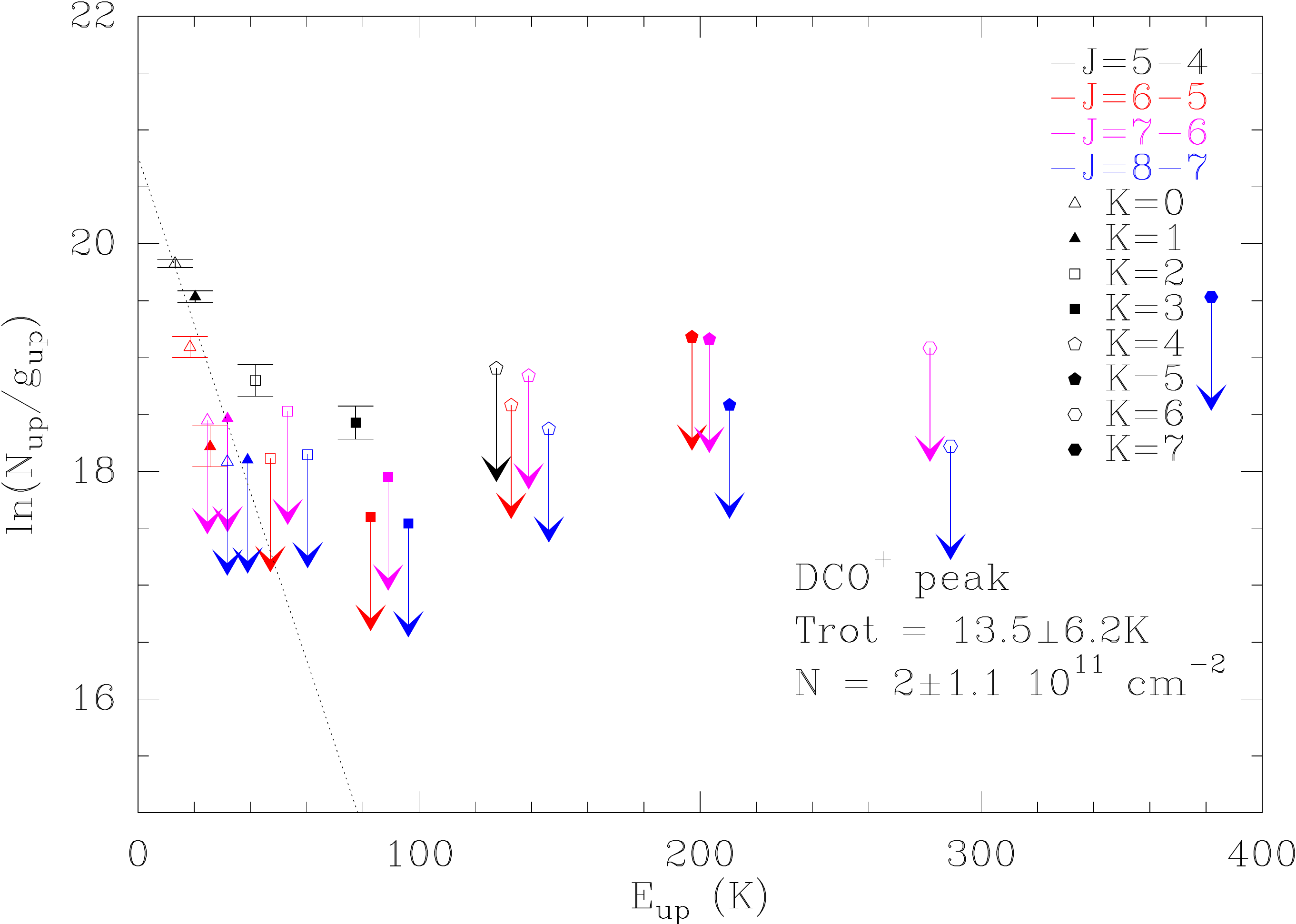}
      \caption{\label{ch3cn_rotdiag_core}{ Same as Fig.~\ref{ch3cn_rotdiag_pdr} for the dense core.}}
    \end{center}
  \end{figure*}}

\newcommand{\FigStacking}{  \begin{figure}
    \begin{center}
      \includegraphics[width=88mm]{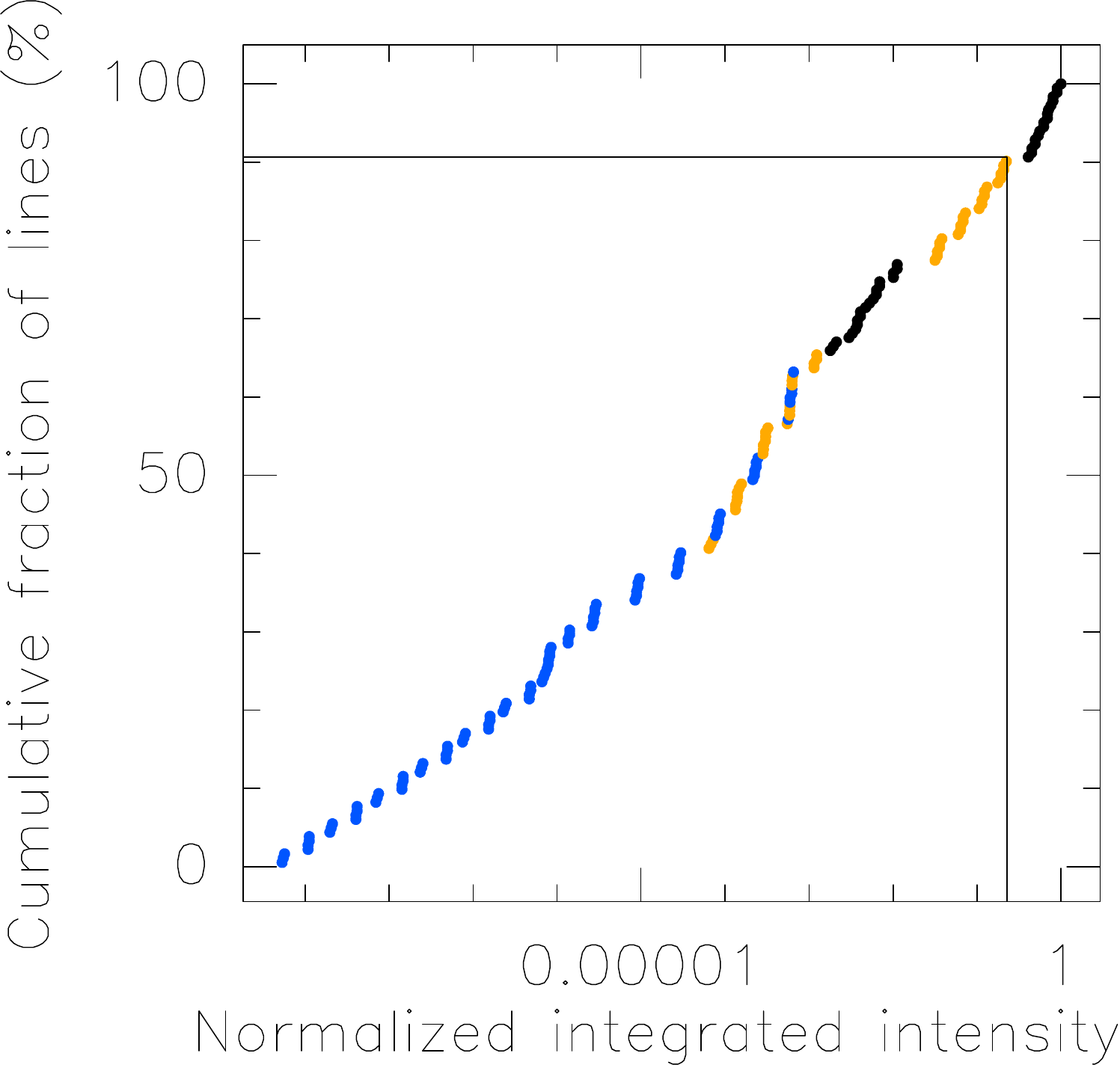}
      \vskip 0.5cm
      \includegraphics[width=88mm]{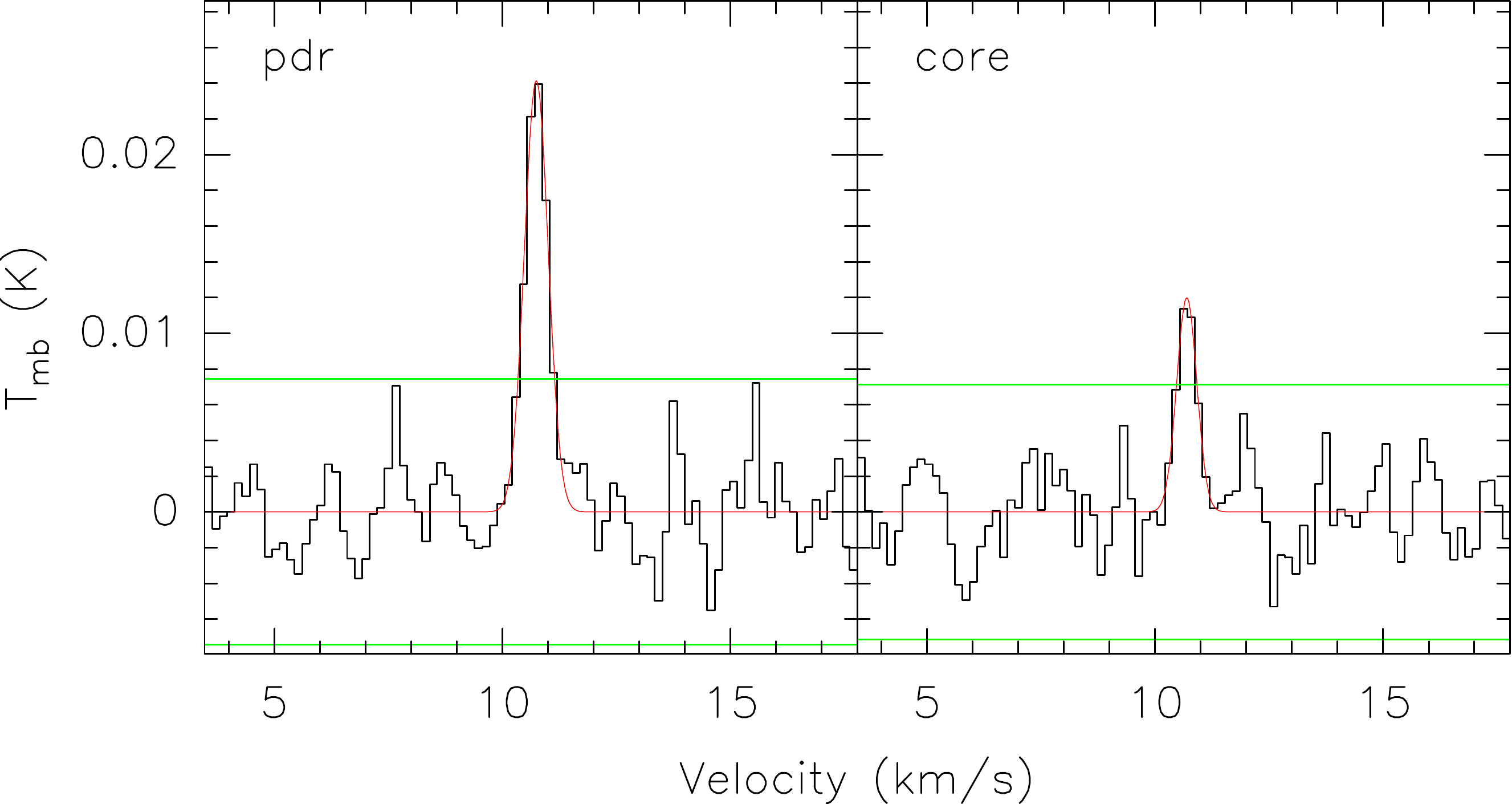}
      \caption{\label{fig.stack}{ \emph{Top:}} Cumulative
        distribution function of computed integrated intensities for
        \chem{C_{3}N} in the optically thin LTE regime with
        $T_\emr{ex}=10\K$, and normalized to the intensity of the brightest
        line. { Only 10\% of all the 1, 2 and 3mm lines have intensities brighter than a fifth
        of the brightest line.  The transitions are color coded by
          bands, black: 3mm, orange: 2mm, blue: 1mm} { \emph{Bottom}}
        \chem{C_{3}N} spectra resulting from the stacking of { 38
          individual lines at 3mm}.  The stacked spectra are shown both for
        the PDR and dense core positions. The red curve is a gaussian fit
        to the stacked spectra, the green lines are $\pm3\sigma$ levels.}
    \end{center}
  \end{figure}}

\newcommand{\TabDipoles}{  \begin{table}
    \begin{center}
    \caption{\label{tab.dipolemom}Dipole moments for the observed species.}
    \begin{tabular}{lrr}
      \hline
      \hline          
      Species & Dipole moment & Reference\\      
              & (Debye)       \\
      \hline
      \metcy     &  3.93    &  \citet{Gadhi.1995}         \\     
      \isometcy  &  3.89    &  \citet{Cernicharo.1988}          \\  
      \cyacet    &  3.73    &   \citet{Deleon.1985}         \\  
      \isocyacet &  2.93    &   \citet{Kruger.1991}          \\  
      \cycethy &  2.85    &   \citet{McCarthy.1995}          \\  
      \hline
      \end{tabular}
     \end{center}
  \end{table}}
  
\newcommand{\TabObservations}{  \begin{table}
    \begin{center}
    \caption{\label{tab.observations}Observation parameters for the observed lines. }
    \begin{tabular}{llrrr}
      \hline
      \hline          
      Molecule & Transition      & Frequency & HPBW  & rms\\      
               &  & (GHz)        &($''$) & (mK [\Tmb{}])\\
      \hline
      \metcy{}     &  J=5--4 K=0,4$^{\dag}$   &       91.972  & 26.9  &  5.8 \\
                   &  J=6--5 K=0,5$^{\dag}$   &      110.356  & 22.4  &  9.8 \\
                   &  J=7--6 K=0,6$^{\dag}$   &      128.734  & 19.2  & 21.8 \\
                   &  J=8--7 K=0,7$^{\dag}$   &      147.105  & 16.8  & 20.9 \\
      \isometcy{}  &  J=5--4 K=0,4$^{\dag}$   &      100.508  & 24.6  &  8.2 \\
      \cyacet{}    &  J=9--8$^{\dag}$         &       81.881  & 30.2  & 19.1 \\
                   &  J=10--9                 &       90.979  & 27.2  &  5.4 \\
                   &  J=11--10                &      100.076  & 24.7  &  8.7 \\
                   &  J=12--11                &      109.173  & 22.7  & 11.0 \\
      \isocyacet{} &  J=9--8                  &       89.419  & 27.7  &  5.4 \\
                   &  J=10--9                 &       99.354  & 24.9  &  7.1 \\
                   &  J=11--10                &      109.289  & 22.6  &  9.3 \\
      \hnccc{}     &  J=9--8                  &       84.028  & 29.4  &  6.0 \\
                   &  J=10--9                 &       93.364  & 26.5  &  5.0 \\
                   &  J=11--10                &      102.700  & 24.1  &  8.1 \\
                   &  J=12--11                &      112.036  & 22.1  & 10.7 \\
      \cycethy{}   &  N=9--8$^{\dag}$         &       89.054  & 27.8  &  6.3 \\
                   &  N=10--9$^{\dag}$        &       98.949  & 25.0  &  8.3 \\
                   &  N=11--10$^{\dag}$       &      108.843  & 22.7  & 11.4 \\
      \hline
  
      \end{tabular}
      \tablefoot{$^{\dag}$The quoted frequencies are characteristic values
        for the multiplet frequencies. The exact frequencies for each line are given in Tabs.~\ref{tab.ch3nc_obs} to \ref{tab.hnccc_obs}.}
     \end{center}
  \end{table}}

\newcommand{\TabMetcyObs}{  \begin{table*}
    \caption{\label{tab.ch3cn_obs}Observed line properties for
      \metcy. }
    \begin{tabular*}{\textwidth}{@{\extracolsep{\fill}} lrrrrrrcccc}
      \hline
      \hline          
      Line & Frequency & \Eu & \gu &    \Au     & \multicolumn{2}{c}{I} & \multicolumn{2}{c}{\Vlsr{}} & \multicolumn{2}{c}{\dV{}}\\      
           & (GHz)       & (K)  &    & (s$^{-1}$) & \multicolumn{2}{c}{($\mK[\Tmb]\,\kms$)}  & \multicolumn{2}{c}{(\kms{})}  & \multicolumn{2}{c}{(\kms{})}\\
      \hline
                  &            &        &     &             & HCO peak       & \DCOp{} peak     & HCO peak  & \DCOp{} peak  & HCO peak      & \DCOp{} peak  \\
      \hline
      J=5--4 K=4  &  91.958726 &  127.5 &  22 &  $2.30(-5)$ &  $ 9.0 \pm6.8$         & $<10 $    &  $10.72\pm0.02$ &  $10.72\pm0.02$   & $0.64\pm0.01$ & $0.74\pm0.03$\\
      J=5--4 K=3  &  91.971130 &   77.5 &  44 &  $4.10(-5)$ &  $ 52.3\pm3.3$ &  $22.1\pm3.5$ &           &            &           &            \\
      J=5--4 K=2  &  91.979994 &   41.8 &  22 &  $5.30(-5)$ &  $ 65.8\pm2.9$ &  $20.7\pm3.1$ &           &            &           &          \\
      J=5--4 K=1  &  91.985314 &   20.4 &  22 &  $6.10(-5)$ &  $125.7\pm2.1$ &  $49.7\pm2.6$ &           &            &          &             \\
      J=5--4 K=0  &  91.987088 &   13.2 &  22 &  $6.30(-5)$ &  $140.1\pm2.3$ &  $68.7\pm2.4$ &           &            &            &             \\
      \hline                                                                                                       
      J=6--5 K=5  & 110.330345 &  197.1 &  26 &  $3.40(-5)$ &  $<12$         & $<16$         &$10.70\pm0.02$ & $10.72\pm0.03$  & $0.55\pm0.02$ & $0.71\pm0.07$ \\ 
      J=6--5 K=4  & 110.349470 &  132.8 &  26 &  $6.20(-5)$ &  $<12$         & $<16$         &           &            &            &           \\
      J=6--5 K=3  & 110.364354 &   82.8 &  52 &  $8.30(-5)$ &  $ 38.4\pm4.4$ & $<16$         &           &            &          &           \\
      J=6--5 K=2  & 110.374989 &   47.1 &  26 &  $9.90(-5)$ &  $ 60.8\pm3.9$ & $<16$         &           &            &          &          \\
      J=6--5 K=1  & 110.381372 &   25.7 &  26 &  $1.08(-4)$ &  $106.4\pm3.8$ &  $29.4\pm3.9$ &           &            &            &           \\
      J=6--5 K=0  & 110.383500 &   18.5 &  26 &  $1.11(-4)$ &  $116.7\pm3.8$ &  $47.7\pm4.6$ &           &            &            &            \\
      \hline   
      J=7--6 K=6  & 128.690111 &  281.8 & 120 &  $4.70(-5)$ & $<28        $& $<34$         &$10.69\pm0.02$ & $\ldots$  & $0.58\pm0.01$ & $\ldots$\\                                                                                        
      J=7--6 K=5  & 128.717359 &  203.3 &  60 &  $8.70(-5)$ & $<28        $& $<34$         &               & $\ldots$            &           &           $\ldots$ \\
      J=7--6 K=4  & 128.739669 &  139.0 &  60 &  $1.20(-4)$ & $<28        $& $<34$         &               & $\ldots$            &            &           $\ldots$ \\
      J=7--6 K=3  & 128.757030 &   89.0 & 120 &  $1.46(-4)$ & $37.2\pm9.2 $& $<34$         &               & $\ldots$            &            &           $\ldots$ \\
      J=7--6 K=2  & 128.769436 &   53.3 &  60 &  $1.64(-4)$ & $26.9\pm8.2 $& $<34$         &               & $\ldots$            &            &           $\ldots$ \\
      J=7--6 K=1  & 128.776882 &   31.9 &  60 &  $1.75(-4)$ & $78.4\pm5.5 $& $<34$         &               & $\ldots$            &            &           $\ldots$ \\
      J=7--6 K=0  & 128.779364 &   24.7 &  60 &  $1.78(-4)$ & $81.4\pm5.7 $& $<34$         &               & $\ldots$            &           &           $\ldots$ \\

      \hline                                                                                                       
      J=8--7 K=7  & 147.035835 &  381.6 &  34 &  $6.30(-5)$ &  $<24$         &  $<31$        & $10.66\pm0.02$& $\ldots$           & $0.54\pm0.05$ &$\ldots$ \\
      J=8--7 K=6  & 147.072602 &  288.8 &  68 &  $1.17(-4)$ &  $<24$         &  $<31$        &               & $\ldots$           &             &  $\ldots$ \\
      J=8--7 K=5  & 147.103738 &  210.3 &  34 &  $1.63(-4)$ &  $<24$         &  $<31$        &               & $\ldots$           &             &         $\ldots$ \\
      J=8--7 K=4  & 147.129230 &  146.1 &  34 &  $2.01(-4)$ &  $<24$         &  $<31$        &               & $\ldots$           &             &          $\ldots$ \\
      J=8--7 K=3  & 147.149068 &   96.1 &  68 &  $2.31(-4)$ &  $21.1\pm7.9$  &  $<31$        &               & $\ldots$           &             &          $\ldots$ \\  
      J=8--7 K=2  & 147.163244 &   60.4 &  34 &  $2.52(-4)$ &  $23.2\pm6.6$  &  $<31$        &               & $\ldots$           &             &          $\ldots$ \\  
      J=8--7 K=1  & 147.171752 &   38.9 &  34 &  $2.64(-4)$ &  $50.5\pm6.8$  &  $<31$        &               & $\ldots$           &             &          $\ldots$ \\
      J=8--7 K=0  & 147.174588 &   31.8 &  34 &  $2.69(-4)$ &  $70.8\pm6.1$  &  $<31$        &               & $\ldots$           &             &         $\ldots$ \\
      \hline
    \end{tabular*}
    \tablefoot{Upper limits at the 95\% (2 sigma) level assuming a 0.7 km/s FWHM}
  \end{table*}}

\newcommand{\TabIsometcyObs}{  \begin{table*}
    \caption{\label{tab.ch3nc_obs}Line properties for \chem{CH_3NC} in
      Tmb. }
    \begin{tabular*}{\textwidth}{@{\extracolsep{\fill}} lrrrrrrcccc}
      \hline
      \hline          
      Line                                            & Frequency                     & \Eu                    & \gu            &       \Au      & \multicolumn{2}{c}{I}                 &       \multicolumn{2}{c}{\Vlsr}        & \multicolumn{2}{c}{$\Delta V$}\\      
           & (GHz)       & (K)  &    & (s$^{-1}$) & \multicolumn{2}{c}{($\mK[\Tmb]\,\kms$)}  & \multicolumn{2}{c}{(\kms{})}  & \multicolumn{2}{c}{(\kms{})}\\

      \hline  
      &                                                         &                              &                      &               & HCO peak             &   \DCOp{} peak  & HCO peak   &   \DCOp{} peak  &  HCO peak       &  \DCOp{} peak   \\
      \hline
      J=5--4 K=4                      & 100.490163      &  127.4 &  22  &$2.92(-5)$  &$<9$    & $<12$ &      $10.71\pm0.026$ & $\ldots$  &       $0.50\pm0.06$ & $\ldots$        \\
      J=5--4 K=3                      & 100.506072      &   78.0 &  44  &$5.20(-5)$  &$12.1\pm2.5$            &  $<12$         &&&&\\
      J=5--4 K=2                      & 100.517433      &   42.7 &  22  &$6.83(-5)$  &$<9$             &  $<12$         &  &&&\\
      J=5--4 K=1                      & 100.524249      &   21.5 &  22  &$7.81(-5)$  &$20.1\pm2.3$      &  $<12$         &                      &&&\\
      J=5--4 K=0                      & 100.526541      &   14.5 &  22  &$8.13(-5)$  &$16.5\pm2.4$      &  $<12$         &                      &&&\\
      \hline
    \end{tabular*}
    \tablefoot{Upper limits at the 95\% (2 sigma) level assuming a 0.7 km/s FWHM}
  \end{table*}}

\newcommand{\TabCyacetObs}{  \begin{table*}
    \caption{\label{tab.hc3n_obs}Line properties for \chem{HC_{3}N} in Tmb.}
    \begin{tabular*}{\textwidth}{@{\extracolsep{\fill}} lrrrrrrcccc}
      \hline
      \hline          
      Line                                            & Frequency                     & \Eu                    & \gu            &       \Au      & \multicolumn{2}{c}{I}                 &       \multicolumn{2}{c}{\Vlsr}        & \multicolumn{2}{c}{$\Delta V$}\\      
           & (GHz)       & (K)  &    & (s$^{-1}$) & \multicolumn{2}{c}{($\mK[\Tmb]\,\kms$)}  & \multicolumn{2}{c}{(\kms{})}  & \multicolumn{2}{c}{(\kms{})}\\
      \hline  
      &                                                         &                              &                      &               & HCO peak             &   \DCOp{} peak  & HCO peak   &   \DCOp{} peak  &  HCO peak       &  \DCOp{} peak   \\
      \hline
      J=9--8                  & 81.881468                     &  19.7 &  17  &$4.2(-5)$ &$86.9\pm8.3$&$113.7\pm7.0$&$10.69\pm0.026$&$10.62\pm0.014$           &       $0.562\pm0.066$ & $0.440\pm0.030$ \\
      J=10--9                         & 90.979023             &  24.0 &  19  &$5.8(-5)$ &$65.6\pm4.0$&$ 95.3\pm3.2$&$10.78\pm0.013$&$10.73\pm0.008$           &       $0.474\pm0.037$ & $0.456\pm0.017$ \\
      J=11--10                        & 100.076392            &  28.8 &  21  &$7.8(-5)$ &$49.6\pm3.3$&$ 69.8\pm4.0$&$10.66\pm0.016$&$10.65\pm0.012$           &       $0.474\pm0.038$ & $0.419\pm0.027$  \\
      J=12--11                        & 109.173634            &  34.1 &  23  &$1.0(-4)$ &$43.2\pm4.5$&$ 55.1\pm4.1$&$10.64\pm0.021$&$10.60\pm0.019$           &       $0.429\pm0.059$ & $0.514\pm0.040$  \\
      \hline
    \end{tabular*}
  \end{table*}}
  
\newcommand{\TabCyethy}{  \begin{table*}
    \caption{\label{tab.c3n_obs}Stacked line properties for \cycethy{}.}
    \begin{tabular*}{\textwidth}{@{\extracolsep{\fill}} lcccc}
      \hline
      \hline          
      Position  &       \Vlsr{}        & $\Delta V$  & SNR\\      
                & (\kms{})  & \kms{}) &    \\              
      \hline  
       HCO peak             &  $10.73\pm0.02$  & $0.629\pm0.055$ & 10.0  \\
       \DCOp{} peak         &  $10.69\pm0.04$  & $0.547\pm0.094$ & 5.1 \\
      \hline
    \end{tabular*}
    \tablefoot{Upper limits at the 95\% (2 sigma) level assuming a 0.7 km/s FWHM}
  \end{table*}}

\newcommand{\TabIsocyacetObs}{  \begin{table*}
    \caption{\label{tab.hc2nc_obs}Line properties for \chem{HC_{2}NC} 
      in Tmb.}
    \begin{tabular*}{\textwidth}{@{\extracolsep{\fill}} lrrrrrrcccc}
      \hline
      \hline          
      Line                                            & Frequency                     & \Eu                    & \gu            &       \Au      & \multicolumn{2}{c}{I}                 &       \multicolumn{2}{c}{\Vlsr}        & \multicolumn{2}{c}{$\Delta V$}\\      
           & (GHz)       & (K)  &    & (s$^{-1}$) & \multicolumn{2}{c}{($\mK[\Tmb]\,\kms$)}  & \multicolumn{2}{c}{(\kms{})}  & \multicolumn{2}{c}{(\kms{})}\\

      \hline  
      &                                                         &                              &                      &               & HCO peak             &   \DCOp{} peak  & HCO peak   &   \DCOp{} peak  &  HCO peak       &  \DCOp{} peak   \\
      \hline
      J=9--8                  & 89.419300                     &  21.5 &  17  & 3.4(-5) &$<9$&$<9$& $\ldots$ &    $\ldots$     &       $\ldots$ & $\ldots$ \\
      J=10--9                         & 99.354250             &  26.2 &   19 &4.7(-5) &$<11$&$<11$& $\ldots$&    $\ldots$     &       $\ldots$ &$\ldots$  \\
      J=11--10                        & 109.289095            &  31.5 & 21   & 6.2(-5)&$<13$&$<13$&$\ldots$ &              $\ldots$&    $\ldots$ &  $\ldots$\\
      \hline
    \end{tabular*}
    \tablefoot{Upper limits at the 95\% (2 sigma) level assuming a 0.7 km/s FWHM}
  \end{table*}}

\newcommand{\TabHNCCCObs}{  \begin{table*}
    \caption{\label{tab.hnccc_obs}Line properties for \hnccc{} 
      in Tmb.}
    \begin{tabular*}{\textwidth}{@{\extracolsep{\fill}} lrrrrrrcccc}
      \hline
      \hline          
      Line                                            & Frequency                     & \Eu                    & \gu            &       \Au      & \multicolumn{2}{c}{I}                 &       \multicolumn{2}{c}{\Vlsr}        & \multicolumn{2}{c}{$\Delta V$}\\      
           & (GHz)       & (K)  &    & (s$^{-1}$) & \multicolumn{2}{c}{($\mK[\Tmb]\,\kms$)}  & \multicolumn{2}{c}{(\kms{})}  & \multicolumn{2}{c}{(\kms{})}\\

      \hline  
      &                                                         &                              &                      &               & HCO peak             &   \DCOp{} peak  & HCO peak   &   \DCOp{} peak  &  HCO peak       &  \DCOp{} peak   \\
      \hline
      J=9--8                 &  84.028469           &  20.2 & 17   & 1.0(-4) &$<9$&$<9$& $\ldots$ &    $\ldots$     &       $\ldots$ & $\ldots$ \\
      J=10--9                & 93.364537            &  24.6 & 19   & 1.4(-4) &$<9$&$<9$& $\ldots$&    $\ldots$     &       $\ldots$ &$\ldots$  \\
      J=11--10               & 102.700471           &  29.6 & 21   & 1.9(-4) &$<12$&$<11$&$\ldots$ &              $\ldots$&    $\ldots$ &  $\ldots$\\
      J=11--10               & 112.036255           &  35.0 & 23   & 2.5(-4) &$<16$&$<16$&$\ldots$ &              $\ldots$&    $\ldots$ &  $\ldots$\\

      \hline
    \end{tabular*}
    \tablefoot{Upper limits at the 95\% (2 sigma) level assuming a 0.7 km/s FWHM}
  \end{table*}}

\newcommand{\TabBayesianPriors}{  \begin{table}
    \caption{Quantitative informations of the
      distribution functions used as priors in the Bayesian fitting.}
    \label{tab.priors}
    \begin{tabular*}{88mm}{@{\extracolsep{\fill}} cccc}
      \hline
      \hline          
      Parameter            &   Type  &                 HCO peak     & \DCOp{} peak \\
      \hline  
      $\log_{10}n_{H_{2}}$ &  normal & $4.8\pm0.2$\tablefootmark{a} & $5.0\pm0.2$\tablefootmark{a} \\ 
      $T_{K}$              &  normal &   $60\pm10$\tablefootmark{b} &   $25\pm10$\tablefootmark{c} \\
      $\log_{10}N$         & uniform &    $[9,17]$                  &   $[9,17]$ \\
      beam dilution        & uniform &     $[0,1]$                  &    $[0,1]$ \\
      \hline
    \end{tabular*}
    \tablefoot{      References:
      \tablefoottext{a}{\citet{Habart.2005}}
      \tablefoottext{b}{\citet{Pety.2005}}
      \tablefoottext{c}{\citet{Goicoechea.2006}}}
  \end{table}}

\newcommand{\TabFitLVG}{  \begin{table*}
    \caption{Best fit parameters for the radiative transfer modeling.}
    \label{tab.fit.lvg}
    \begin{tabular}{llccc|cccc}
      \hline
      \hline
      Position & Species   & Smallest observed beam &$\log_{10}N_{H_{tot}}$& $[e^{-}]=n_{e}/n_{\Ht}$ & $T_\emr{K}$ & $\log_{10}n_{H_{2}}$ & $\log_{10}N$ & $\log_{10}X$\tablefootmark{a} \\
               &         &  ($''$)                        &      (\pscm{})      &            &  (K) &  (\pccm{})    &(\pscm{})                &                             \\
      \hline  
      HCO peak     & \metcy{}  & 16.8  &$22.6$  & 0          & $43\pm1$ & $4.3\pm0.1$ & $13.2\pm0.2$  &  $ -9.4\pm0.2$ \\
                   & \metcy{}  & 16.8  &$22.6$  & $10^{-4}$  & $45\pm2$ & $4.3\pm0.3$ & $13.0\pm0.2$  &  $ -9.6\pm0.2$ \\
                   & \cyacet{} & 22.7  &$22.6$  & 0         & $60\pm11$ & $4.7\pm0.2$ & $11.5\pm0.1$ &  $-11.1\pm0.1$ \\
                   & \cyacet{} & 22.7  &$22.6$  & $10^{-4}$ & $58\pm11$ & $4.6\pm0.2$ & $11.4\pm0.1$ &  $-11.2\pm0.1$ \\
      \hline    
      \DCOp{} peak & \metcy{}  & 16.8  &$22.8$  & 0         & $33\pm3$ & $5.0\pm0.2$ & $11.7\pm0.2$  &  $-11.1 \pm0.2$\\
                   & \cyacet{} & 22.7  &$22.8$  & 0         & $30\pm6$ & $4.8\pm0.2$ & $11.7\pm0.2$  &  $-11.1 \pm0.2$\\
      \hline  
    \end{tabular}
    \tablefoot{ { The left hand side of the table corresponds to fixed
      parameters, and the right hand side to computed values.}\\
      \tablefoottext{a}{$X = \frac{N(X)}{N(H)+2N(H_{2})}.$}}
  \end{table*}}

\newcommand{\TabCCCN}{  \begin{table}
    \begin{center}
    \caption{\label{tab.cccn} Column density and abundances for \cycethy{}.}
    \begin{tabular}{lrr}
        \hline
        \hline          
        Position & Column density & Abundance \\      
                 & (\pscm{})      &           \\
        \hline
        PDR      & $\sciexp{(2\pm1)}{12}$  & $\sciexp{(5\pm2.5)}{-11}$ \\
        Core     & $<\sciexp{1}{12}$       & $<\sciexp{1.5}{-11}$       \\
        \hline
      \end{tabular}
     \end{center}
  \end{table}}

\newcommand{\TabLineRatios}{  \begin{table}
    \caption{\label{CH3NC_CH3CN_lineratio}\isometcy{}/\metcy{} line
      ratio for different K levels of J=5-4. } 
    \begin{center}
      \begin{tabular}{lrr}
        \hline
        \hline
        Line & HCO peak  & \DCOp{} peak\\
        \hline
        K=0 & $0.12\pm0.02$       &       $<0.17$\\
        K=1 & $0.16\pm0.02$       &       $<0.24$\\
        K=2 & $<0.14$    &       $<0.57$\\
        K=3 & $0.23\pm0.05$       &       $<0.54$\\
        K=4 & $<1$    &       $<1.2$\\
        \hline
      \end{tabular}      
    \end{center}
  \end{table}}
  
\newcommand{\TabBeamDilution}{  \begin{table}
    \caption{\label{tab.beam}Beam dilution for \metcy{}
        determined from the radiative transfer modeling. } 
     \begin{center}
      \begin{tabular}{lllcc}
      \hline
      Species & Line & Resolution &  \multicolumn{2}{c}{Dilution ($2\sigma$ confidence interval)}\\
              &      &    ($''$)  &  PDR  & dense core\\
      \hline
      \metcy{}  & J=5-4   & 26.9 & $[0.19-0.31]$ & $[0.65-1.00]$\\
                & J=6-5   & 22.4 & $[0.31-0.45]$ & $[0.52-0.88]$\\
                & J=7-6   & 19.2 & $[0.49-0.63]$ &    ---       \\
                & J=8-7   & 16.8 & $[0.82-1.00]$ &    ---       \\
      \hline
       \end{tabular}      
    \end{center}
  \end{table}}

\begin{document}

\title{The IRAM-30m line survey of the Horsehead PDR:\\
  III. High abundance of complex (iso-)nitrile molecules 
  in UV-illuminated gas\thanks{Based on observations obtained with the
    IRAM-30m telescope.  IRAM is supported by INSU/CNRS (France), MPG
    (Germany), and IGN (Spain).}}

\titlerunning{High abundance of complex (iso-)nitrile molecules in the
  Horsehead PDR}

\author{P.~Gratier \inst{\ref{IRAM}} \and
  J.~Pety\inst{\ref{IRAM},\ref{LERMA}} \and V. Guzm\'a{}n\inst{\ref{IRAM}}
  \and M. Gerin\inst{\ref{LERMA}} \and J. R. Goicoechea\inst{\ref{CAB}} \and
  E. Roueff\inst{\ref{LUTH}} \and A. Faure \inst{\ref{IPAG}}}

\authorrunning{Gratier P., Pety J., Guzm\'a{}n V. \etal{}}

\institute{  Institut de Radioastronomie Millim\'etrique, 300 rue de la Piscine, 38406 Saint Martin d'H\`eres, France\\
  \email{[gratier,pety]@iram.fr}\label{IRAM}  \and LERMA, UMR 8112, CNRS and Observatoire de Paris, 61 avenue de
  l'Observatoire, 75014 Paris, France\label{LERMA}   \and Centro de Astrobiolog\'ia. CSIC-INTA. Carretera de Ajalvir, Km 4.
  Torrej\'on de Ardoz, 28850 Madrid, Spain\label{CAB}   \and LUTH UMR 8102, CNRS and Observatoire de Paris, Place J. Janssen,
  92195 Meudon Cedex, France\label{LUTH}   \and UJF-Grenoble 1/CNRS-INSU, Institut de Plan\'etologie et
  d'Astrophysique de Grenoble (IPAG) UMR 5274, 38041 Grenoble,
  France\label{IPAG}}

\date{}

\abstract
{Complex (iso-)nitrile molecules, such as \metcy{} and \cyacet{}, are relatively
easily detected in our Galaxy and in other galaxies.}	
{We aim at constraining their chemistry through observations of two positions in
the Horsehead edge: the photo-dissociation region (PDR) and the dense, cold, and
UV-shielded core just behind it.}
{We systematically searched for lines of \metcy{}, \cyacet{}, \cycethy{}, and
some of their isomers in our sensitive unbiased line survey at 3, 2, and 1\mm{}.
We stacked the lines of \cycethy{} to improve the detectability of this species.
We derived column densities and abundances through Bayesian analysis using a
large velocity gradient radiative transfer model.}
{We report the first clear detection of \isometcy{} at millimeter wavelength. We
detected 17 lines of \metcy{} at the PDR and 6 at the dense core position, and we
resolved its hyperfine structure for 3 lines. We detected 4 lines of \cyacet{},
and \cycethy{} is clearly detected at the PDR position. We computed new electron
collisional rate coefficients for \metcy{}, and { we found that including
electron excitation reduces the derived column density by 40\% at the PDR
position, where the electron density is 1--5\pccm{}.} While \metcy{} is 30 times
more abundant in the PDR (\sciexp{2.5}{-10}) than in the dense core
(\sciexp{8}{-12}), \cyacet{} has similar abundance at both positions
(\sciexp{8}{-12}). The isomeric ratio \isometcy{}/\metcy{} is $0.15\pm0.02$.}
{The significant amount of complex (iso-)nitrile molecule in the UV illuminated
gas is puzzling as the photodissociation is expected to be efficient. This is all
the more surprising in the case of \metcy{}, which is 30 times more abundant in
the PDR than in the dense core. In this case, pure gas phase chemistry cannot
reproduce the amount of \metcy{} observed in the UV-illuminated gas. We propose
that \metcy{} gas phase abundance is enhanced when ice mantles of grains are
destroyed through photo-desorption or thermal-evaporation in PDRs, and through
sputtering in shocks.}
   
\maketitle{}    
\section{\label{sec.intro}Introduction}

Complex nitriles like \metcy{} and \cyacet{} are easily detected in (massive)
star-forming regions~\citep{Araya.2005,Bottinelli.2004,Purcell.2006}. { \metcy{}
and \cyacet{} were detected in the Mon~R2 ultracompact \ion{H}{ii}
region~\citep[][]{Ginard.2012}.} \citet{Mauersberger.1991} reported the first
detection of \metcy{} in M\,82 and NGC\,253. \citet{Lindberg.2011} detected
\cyacet{} in 13 local universe galaxies. { These molecules} are often used to
constrain the physical conditions of the host gas. In particular, \metcy{} is
thought to be a good thermometer because it exhibits sets of metastable
transitions, which are only coupled through collisions, but not
radiatively~\citep[]{Guesten.1985}.

Moreover, \metcy{} is easily detected towards hot molecular
cores~\citep{Olmi.1996,Olmi.1996a,Hatchell.1998,Purcell.2006}. In particular,
\citet{Purcell.2006} detect 3mm lines of \metcy{} in 58 candidate hot molecular
cores on a sample of 83 methanol maser-selected star-forming regions. They detect
\metcy{} in isolated methanol maser sites and find that \metcy{} is more
prevalent and brighter when an ultracompact \ion{H}{ii} region is present,
independent of the distance to the source. The inferred \metcy{} abundances are
higher than can be accounted for by current pure gas phase chemical
modeling~\citep[\eg{}][]{Olmi.1996,Hatchell.1998}. \citet{Mackay.1999} proposed
that the \metcy{} abundance could be explained if the abundances of the gas phase
precursors of \metcy{}, \ie{} \chem{CH_{3}^{+}} and HCN, are enhanced by
evaporation from grain ices and by further photo-processing of methanol and
ammonia.

The mane of the Horsehead nebula is an ideal source to test the excitation and
chemistry of nitriles. Indeed, it is viewed nearly edge-on \citep{Abergel.2003}
at a distance of 400\pc{} (implying that $10''$ correspond to 0.02\pc{}). It has
a steep density gradient, from $\nH\sim100\pccm$ in the UV illuminated outer
layers rising to $\nH\sim2\times10^5\pccm$ in less than $10''$
\citep{Habart.2005}. \citet{Gerin.2009a} show that the HCO emission delineates
the UV illuminated edge of the nebula. Less than $40''$ away from the HCO peak
emission where the gas is warm ($\Tkin \sim 60$~K), there is a UV-shielded,
dense, and cold condensation ($\Tkin \leq 20$~K), where HCO$^+$ is highly
deuterated \citep{Pety.2007}. For simplicity, the HCO and \DCOp{} emission peaks
are hereafter referred to as the PDR and dense core positions. The moderate
illumination~\citep[$\chi \sim 60$][]{Abergel.2003,Draine.1978} translates into
dust temperatures low enough that thermal evaporation of complex molecules is
negligible. In the Horsehead nebula one can thus isolate photo-desorption effects
for complex molecules like formaldehyde, which can form on the grain ice mantles
before being photo-desorbed into gas phase, as was shown by \citet{Guzman.2011}.

\FigHorseheadOverview{}   
In this work, we report the detection of \metcy{}, \isometcy{}, \cyacet{}, and
\cycethy{} in the PDR and/or the dense core positions of the Horsehead edge.
Determining the isomeric abundance ratio is an important tool for constraining
the chemical routes to form the molecules. Indeed, the observed column density
ratios often differ significantly from the ones expected by the energies of the
molecules at thermodynamic equilibrium. One such example related to the
\metcy{}/\isometcy{} ratio is the ratio between HNC and HCN which is found in the
interstellar medium to be close to 1 \citep{Sarrasin.2010,Mendes.2012}, even
though HNC is less stable than HCN by 0.5~eV. This gives constraints on the
mechanisms of formation and destruction of these species. We thus also searched
for emission of \isometcy{} and \isocyacet{}. Table~\ref{tab.dipolemom} shows
that all these species have similar dipole moments.

Sect.~\ref{sec.obs} presents the observations and the spectroscopy of the
observed molecules. Sect.~\ref{sec.results} summarizes the observed line
properties. Sect.~\ref{sec.abundances} explains how the column densities are
computed and it presents the inferred abundances. The implications of these
results for the chemistry of \metcy{} and \cyacet{} are discussed in
Sect.~\ref{sec.discussion}. Appendix~\ref{app.obs} summarizes the line properties
(integrated intensity, line width, systemic velocity) for observed lines and
derived upper limits for undetected lines. Appendix~\ref{app.mcmcradex} details
the Bayesian approach used to infer the abundances from the line brightnesses.
Finally, Appendix~\ref{app.rot} discusses the shortcoming of the rotational
diagram method in the subthermal excitation conditions found in the PDR
position. Larger versions of
Figs.~\ref{fig.ch3nc_spec}~and~\ref{fig.ch3cn_spec} are presented in
Appendix~\ref{app.largefig}. \section{\label{sec.obs}Observations and
spectroscopy}

\TabDipoles{} 
After a short description of the Horsehead WHISPER unbiased line survey, this
section presents the spectroscopy of the different molecules studied here,
including their hyperfine structure.

\subsection{Horsehead WHISPER: An unbiased line survey}

The data\footnote{Published WHISPER data are available on the project website:
\url{http://www.iram.fr/~horsehead/}} presented in this paper are extracted from
the Horsehead WHISPER project (Wideband High-resolution Iram-30m Surveys at two
Positions with Emir Receivers, PI: J.~Pety), an unbiased line survey of the 3, 2
and 1mm band, which is currently completed with the IRAM-30m telescope.
This project was observed in 2011 and 2012. Two positions are observed:
1) the HCO peak that is characteristics of the photo-dissociation region at the
surface of the Horsehead nebula~\citep{Gerin.2009a}, and 2) the \DCOp{} peak that
belongs to the nearby cold dense core at high visual
extinction~\citep{Pety.2007}. The combination of the new EMIR receivers and the
Fourier Transform Spectrometers yield a spectral survey with unprecedented
bandwidth (36GHz at 3mm, 34GHz at 2mm, and 76GHz at 1mm), spectral resolution
(49kHz at 3 and 2mm and 195kHz at 1mm, this corresponds a velocity
resolution between 0.1 and 0.3\kms{}), and sensitivity (median noises of 8.0mK,
18.5mK, and 8.3mK respectively). This allowed us to detect $\sim150$ lines from
$\sim30$ species and their isotopologues. { Each sky frequency was observed with
two different frequency tunings. The Horsehead PDR and dense core positions (see
Fig.~\ref{fig.horsehead_overview}) were alternatively observed every 15 minutes
in position switching mode with a common fixed off position (offset: $-100''$,
$0''$ from \radec{05}{40}{54.27}{-02}{28}{00}). The total observing time amounted
to one hour per frequency setup and position. This observing strategy allows us
to remove potential ghost lines resulting from the incomplete attenuation of
strong spectral features in the image side band (the typical rejection of the
EMIR sideband separating mixers is 13dB or a factor 20). A detailed presentation
of the observing strategy and data reduction process will be given in another
paper. Tab.~\ref{tab.observations} summarizes the beamwidths and noise levels for
the frequency ranges of the survey corresponding to the lines discussed in this
paper.}

\TabObservations{}

\subsection{\metcy{} \& \isometcy{}: Two symmetric top species}

\FigMetcyEdiagram{} 
\subsubsection{Energy diagrams}

Fig.~\ref{fig.ch3cn_energy} displays the \metcy{} energy diagram. The \isometcy{}
energy diagram has a similar structure. Indeed, as prolate symmetric top
molecules, their rotational energy (\EJK{}) is derived \citep{Gordy.1984} from
the moments of inertia and angular momentum through

\begin{eqnarray*}
  \frac{\EJK}{h} &= \Be \Jq (\Jq+1) + (\Ae-\Be)\Kq^2 - \D{\Kq}\Kq^4 \\
                 &  - \D{\Jq} \Jq^2 (\Jq+1)^2 - \D{\Jq,\Kq} \Jq (\Jq+1)\Kq^2,
\end{eqnarray*}

where 1) the \Jq{} and \Kq{} quantum numbers describe respectively the total
angular momentum and its projection along the molecule symmetry axis, 2) \Ae{}
and \Be{} are the rotational constants, characteristics of the moments of inertia
along and perpendicular to the symmetry axes, and 3) the \D{\Jq{}}, \D{\Kq{}} and
\D{\Jq{},\Kq{}} are the centrifugal stretching constants. As a symmetric top, the
degeneracy on the symmetry axis is removed, giving a splitting of each rotational
J level into doubly degenerate K components, with $K = 0, \ldots, J$ (except when
$K=0$).

\subsubsection{Transitions}

Using the rules for allowed rotational transitions \citep{Gordy.1984}
\begin{equation}
  \Delta K = 0 \mbox{ and } \Delta J = \pm 1,
  \label{eq:transitions}
\end{equation}
the frequencies of a (\Jq{},\Kq{}) transition can then be derived as
\begin{eqnarray*}
  \nu_{(\JKsimple{J+1}{J}{K})} = 2 \Be (\Jq+1)  - 4 \D{\Jq} (\Jq+1)^3 - 2\D{\Jq,\Kq} (\Jq+1)\Kq^2.
\end{eqnarray*}

A set of ($\Jo{J+1}{J}$, $\Kq=\emr{cst}$) lines is named a \Jq{}-ladder, and a
set of ($\Jq=\emr{cst}$, $\Ko{K+1}{K}$) lines is named a \Kq{}-ladder.

For a given \Jq{}, the probed energy range is mainly determined by
$(\Ae-\Be)\Kq^2$ (a first order effect), while the frequencies are separated by
$2\D{\Jq,\Kq} (\Jq+1)\Kq^2$ (a second order effect due to the centrifugal
distortion). Hence, each \Jq{}-ladder can be observed simultaneously in a
relatively narrow frequency band, while they probe a wide range of energies,
typically from 10 to a few 100\K{}. This is why the excitation of symmetric top
molecules are good thermometers when the lines are thermalized.

\FigIsometcySpectra{}\FigMetcySpectra{} 

\subsection{\cyacet{}, \isocyacet{}, \hnccc{} \& \cycethy{}: Four rigid rotors}
  
\cyacet{}, \isocyacet{} and \hnccc{} are linear species with a $^{1}\Sigma^+$
electronic state. Their spectroscopy is thus much simpler than the symmetric top
ones. Their rotational spectrum is well described by the rigid rotor
approximation with lines separated by 9.1~GHz, 9.9~GHz, 9.3~GHz respectively.
\isocyacet{} and \hnccc{} are isomers of \cyacet{}, which can also be described
as rigid rotors. \cycethy{} has a more complex $^{2}\Sigma$ ground state
electronic structure, which exhibits doublets of nearby frequencies.

\subsection{Hyperfine splitting}

All of the observed N species exhibit a hyperfine structure, {although that of
the $\chem{^{15}N}$ species are not resolved due to the small magnetic dipole
coupling term.} The NC isomers have lower hyperfine splitting values than the
CN isomers. This comes from the fact that the electric field gradient, which
creates the hyperfine splitting through interaction with the nuclear electric
quadrupole moment of $^{14}$N, is stronger for outermost N positions. This is
well known for HCN and HNC \citep[see][]{Bechtel.2006}.

As the Horsehead PDR is seen edge-on, the lines are narrow with typical full
width at half maximum values of 0.6-0.8\kms{}. This enables us to resolve the
hyperfine splitting when it is large enough. In the case of symmetric top
molecules, the hyperfine splitting increases with increasing \Kq{} levels and
decreasing \Jq{} level. In practice, the hyperfine splitting is resolved only for
the (\JK{5}{4}{2}), (\JK{5}{4}{3}), and (\JK{6}{5}{3}) lines of \metcy{} in our
observations (see Fig.~\ref{fig.ch3cn_spec}).

We wish to estimate the correct integrated intensities for each ($\Delta J,K$)
line even though hyperfine components can be blended. To do this, we fitted
together multiple Gaussian profiles over all of the hyperfine levels for each
($\Delta \Jq=1, \Kq = \emr{cst}$) transition. In this global fit, we fixed 1) the
relative intensities of the hyperfine levels to the optically thin values and 2)
the frequency offsets of all the lines to the CDMS catalog values
\citep{Muller.2005a}. There remains $\Jq+2$ free parameters, namely a global
linewidth, a global velocity shift and a multiplicative amplitude factor for each
of the hyperfine set of lines in the ($\Delta \Jq=1, \Kq = \emr{cst}$), with
$\Kq=0,...,\Jq-1$. The integrated intensity of each $(\Delta J,K)$ transition is
then obtained by summing the integrated intensities of the individual Gaussian
functions fitted to the hyperfine structure. This method is only correct for
optically thin lines. Sect.~\ref{sec.opacity} shows that this is case for the
spectral lines we analyze in this paper.

\section{Observational results}
\label{sec.results}

\subsection{First clear detection of \isometcy{} in the millimeter domain}

Figure~\ref{fig.ch3nc_spec} shows the 3 lines of \isometcy{} detected in the
Horsehead PDR. They belong to the 5 lines of the $J=5-4$ \Kq{}-ladder of
\isometcy{}. None of them were detected in the UV-shielded dense core. As the
hyperfine splitting is not resolved for any of these lines, we simultaneously
fitted a single Gaussian profile at each frequency found in the JPL
database~\citep{Pickett.1998}. Moreover, we required a common linewidth for the
different lines. The fit results are displayed in Table~\ref{tab.ch3nc_obs}. The
line intensity modeling carried out in Sect.~\ref{sec.abundances} predicts line
intensities for the other \isometcy{} lines in the 1, 2 and 3\mm{} bands well
below our detection limit.

This is the first clear detection of this molecule in the millimeter domain. A
detection of the $J=1-0$ line of \isometcy{} at centimeter wavelengths (\ie{},
20.1~GHz) has been reported by \citet{Irvine.1984} in TMC-1 and
\citet{Remijan.2005} in SgrB2. At millimeter wavelengths, \citet{Cernicharo.1988}
reported a tentative detection of the $J=4-3$, $J=5-4$ and $J=7-6$ lines in SgrB2
but the large linewidths (20\kms{} FWHM) and contamination from numerous other
lines prevented a robust identification.

\subsection{\metcy{} lines are brighter in the PDR than in the dense core}

In our line survey, four $\Jo{J+1}{J}$ \Kq{}-ladders are detected: The
$J=5\rightarrow4$, $J=6\rightarrow5$ \Kq{}-ladders in the 3mm atmospheric window
at 91.9 GHz and 110.3 GHz respectively and, the $J=7\rightarrow6$ and
$J=8\rightarrow7$ \Kq{}-ladders in the 2mm window at 128.7 GHz and 147.1 GHz.
Figure~\ref{fig.ch3cn_spec} displays the 17 and 6 detected lines at the PDR and
dense core positions, respectively. Lines from the additional 5 $\Jo{J+1}{J}$
\Kq{}-ladders (from $J=11-10$ to $J=15-14$), whose wavelengths lie at 1mm, remain
undetected in our survey. The corresponding noise levels for these
undetected lines are 15--30\mK\ per 200\kHz{} channel.

The derived line properties are synthesized in Table~\ref{tab.ch3cn_obs}. Not
only are there many more detected lines of \metcy{} in the PDR than in the dense
core, but the detected lines are also brighter in the less dense, UV-illuminated PDR
position. Moreover, when the lines are detected in both environment, they
systematically show a narrower linewidth in the PDR than in the core.

\subsection{\isometcy{}/\metcy{} isomeric line ratios}

\TabLineRatios{} 
The value of the ratios of the \isometcy{}/\metcy{} integrated intensities for
the $\Jq=5-4$ \Kq{}-ladder are given in Table~\ref{CH3NC_CH3CN_lineratio} for
both the PDR and dense core positions. The weighted average line ratio is
$0.15\pm0.02$ at the PDR position. The individual ratios exhibit a small scatter
around this value.

\subsection{\cyacet{} lines have similar brigthnesses in the PDR and the
  dense core}

\FigCyanoacetSpectra{}
Four \cyacet{} lines lie in the 3mm, 2 in the 2mm and 8 in the 1mm band we have
observed. All four 3mm lines of \cyacet{} are detected at both observed
positions, no detection were obtained at 2 and 1mm. Independent Gaussian profiles
were fitted for each detected line. Fig.~\ref{fig.hc3n_spec} displays the
\cyacet{} lines, which are detected in our survey and Table~\ref{tab.hc3n_obs}
summarizes the fit results. In contrast with the results for \metcy{}, \cyacet{}
lines are slightly more luminous in the UV-shielded dense core.

\isocyacet{} has three lines in the 3mm band. None of these are detected in any
of the two observed positions. \hnccc{} has four lines in the 3mm band, none of
them are detected. Table~\ref{tab.hc2nc_obs} gives the derived upper limits for
both species. Stacking with the method presented in Sect.~\ref{c3n_obs}
did not reveal any detection. For reference, \isocyacet{} and \hnccc{} have both
been detected in dark clouds \citep{Kawaguchi.1992,Kawaguchi.1992a}, while only
\isocyacet{} has been detected in circumstellar envelopes \citep{Gensheimer.1997}.

\subsection{\label{c3n_obs}\cycethy{} is twice as bright in the PDR than in the
dense core}

Because of its $^{2}\Sigma$ ground state electronic structure, and of the
hyperfine splitting due to the Nitrogen atom, the energy radiated by \cycethy{}
is spread over a large number of lines. This implies that individual lines of
\cycethy{} are less easily detected than, \eg{}, that of \metcy{} in the same
conditions of noise. We thus have coadded the spectral regions where individual
lines from \cycethy{} were expected to show up. This method can be applied to any
species expected to present a rich spectrum with numerous weak spectral lines of
similar intensity.

Using the list of transitions observable in our survey from public catalogs, we
averaged the intensities of each potential line after 1) aligning each spectral
window to the same LSR velocity, and 2) resampling each line spectrum to the same
velocity resolution. In this process, we reject all partss of the original
spectrum which could be contaminated by a line from another already detected
species. We first tried a simple noise-weighted average, \ie{},

\begin{equation}
  S(v) = \frac{\sum_i \frac{T_i(v)}{\sigma_i^2}}{\sum_i \frac{1}{\sigma_i^2}},
\end{equation}

where $v$ is the velocity, $T_i(v)$, and $\sigma_i$ the brightness temperature
and noise of the \emph{i}-th transition, and $S(v)$ the stacked spectra. This is
however a too simple approach because the searched species have complex hyperfine
structures with expected line intensities which vary over several orders of
magnitude. Hence, the signal can easily be drowned in noise. It is thus important
to also weight the different lines according to their expected integrated
intensities, $W_i$,
	
\begin{equation}
  S(v) = \frac{\sum_i \frac{W_i}{\sigma_i^2}\,T_i(v)}{\sum_i \frac{W_i}{\sigma_i^2}}
       = \frac{\sum_i \frac{w_i}{\sigma_i^2}\,T_i(v)}{\sum_i \frac{w_i}{\sigma_i^2}}
  \quad \mbox{with} \quad
  w_{i}=\frac{W_{i}}{\Sigma W_{i}}.
\end{equation}

The relative integrated intensities, $w_i$, are computed with a simple local
thermodynamic equilibrium (LTE) approach. In the optically thin regime, only
the excitation temperature fixes the $w_i$, because the normalization cancels
the contribution of the total column density.

\FigCCCNSpectraLTE{} \FigStacking{} 

We explored a range of excitation temperatures and we kept the stacked spectra
corresponding to the highest derived peak temperature. This strategy was checked
against a spectrum simulated with a LTE code, assuming a typical excitation
temperature of 10\K{}, consistent with our previous studies of complex molecules
in the Horsehead edge, which showed that they are subthermally
excited~\citep{Guzman.2011,Pety.2012}. The proposed strategy recovered the right
excitation temperature.

There are 7+4+3 \cycethy{} doublets in the 1, 2, and 3\mm{} bands, each one split
by hyperfine interaction into a total of 211 lines. Observationally, this results
into six marginal detections (\ie{}, a peak signal-to-noise ratio lower or close
to 3) at the PDR position (see Fig.~\ref{fig.c3n_lte}).

{ Fig.~\ref{fig.stack} shows 1) the distribution of \chem{C_{3}N} line
brightnesses modeled assuming optically thin LTE emission with $T_\emr{ex}=10\K$,
and 2) the \cycethy{} spectra { obtained by stacking the 38 3mm lines} at the PDR
and dense core position. The contributions from all 173 lines at 1\mm{} and
2\mm{} is negligible in our case, as none of them are brighter than 20\% of the
brightest 3mm band line. Including the 1\mm{} and 2\mm{} lines would have reduced
the spectral resolution of the stacked spectrum as individual spectra must first
be resampled to the coarsest velocity resolution ($0.282\kms{}$).}

\cycethy{} is twice as bright in the PDR than in the dense core. The brightest
\cycethy{} line corresponds to a telescope half primary beam width of $27.7''$,
the proportion of the line intensity observed towards the dense core that arises
from beam pickup from the PDR position is less than 8\%. The remaining
emission could arise in the lower density skin of the dense core, already
detected in HCO~\citep{Gerin.2009a} and
\chem{CF^{+}}~\citep{Guzman.2012,Guzman.2012a}.

\section{\label{sec.abundances}Column densities and abundances}

\subsection{Tools}

\subsubsection{Local thermodynamic equilibrium vs escape probability
  radiative transfer}

Detailed excitation and radiative transfer calculations are needed to estimate
the line intensities of interstellar species from the source physical properties
(gas density, temperature and source size) and the species column density. The
inverse problem of deriving physical conditions and column densities for observed
line intensities needs assumptions. For instance, the rotation diagram analysis
can be used when the populations of the energy levels are in Local Thermodynamic
Equilibrium~\citep[see][for a detailed description of non LTE and optical
thickness effects on the rotational diagram method]{Goldsmith.1999}. The critical
densities for the methylcyanide lines are typically $10^{5}-10^{6}\pccm$, while
the typical gas density and temperatures are $6\times10^4\pccm$ and $60\K$ in the
Horsehead PDR, and $1\times10^5\pccm$ and $25\K$ in the dense core.

The lines detected in the Horsehead are thus subthermally excited. In this case,
both collisional and radiative (de)excitation must be taken into account. For
instance, escape probability
methods~\citep{Sobolev.1960,Castor.1970,Goldreich.1974} correctly treat the
radiative transfer micro-physics. However, they suppose the presence of a large
velocity gradient so that photons escape their local environment, \ie{},
radiative transfer is only local. RADEX \citep{van-der-Tak.2007} is one such
method. We use it here and we compare the results with those obtained
in LTE.

\subsubsection{\label{sec.electron}Hydrogen vs electron excitation}

In order to correctly treat the micro-physics, radiative transfer methods need
collisional (de)excitation coefficients as inputs. In general, only collisions
with the most abundant gas species, \ie{} ortho and para \Ht{} and helium, are
taken into account. However, excitation by electrons are expected to play a
significant role when the electron fraction reaches
$[\emr{e}^{-}]=n_{e}/n_{\Ht}\sim 10^{-5}-10^{-4}$. The importance of the
collisional exception with these two families of partners are similar when
$n_\emr{e} C_\emr{e} \sim n_{\Ht{}} C_{\Ht{}}$. While collisional coefficients
with neutral are typically $10^{-11}-10^{-10} \ccm\ps{}$, those with electrons
are around $10^{-6}$. \citet{Goicoechea.2009a} determined an electron fraction of
$10^{-4}$ at the PDR position, implying that electrons contribute significantly
to the excitation. In contrast, the same study derived an electron fraction of a
few $10^{-9}$ in the dense core, where the electron excitation is thus negligible.

We have computed the CH$_3$CN-e$^-$ and HC$_3$N-e$^-$ collisional coefficients
within the dipolar Born approximation \citep[\eg{}][]{Itikawa.1971}. Owing to the
large dipole of both species (3.73~D for HC$_3$N and 3.92~D for CH$_3$CN),
dipole-allowed cross sections are indeed expected to be dominant and entirely
determined by the long-range electron-dipole interaction \citep[see
\eg{}][]{Faure.2007}. In this approximation, cross sections are proportional to
line strengths (and the square of the dipole) and therefore strictly obey the
dipolar selection rule (see Eq.~\ref{eq:transitions}). Line strengths and dipoles
were taken from the CDMS catalog \citep{Muller.2005a}. Excitation cross sections
were computed in the energy range 0.1\,meV$-$1\,eV and rate coefficients were
deduced in the range 10-1000\K{}, for the lowest 251 levels of CH$_3$CN (238
transitions) and the lowest 31 levels of HC$_3$N (30 transitions). The
CH$_3$CN--e$^-$ and HC$_3$N--e$^-$ collisional rates are available online through
the BASECOL\footnote{\url{http://basecol.obspm.fr}} database.

\subsubsection{Beam dilution}

Beam dilution arises when the source does not fill the beam. In our observations,
the half primary beam width of the IRAM-30m varies from $30''$ at the low end of
the 3mm band to $10''$ at the high end of 1mm band. Moreover, a PDR lies by
definition at the interface between fully ionized and molecular gas. Previous PDR
modeling of the Horsehead edge implies that the physical and chemical typical
angular scales ranges from 1 to $50''$. It thus is likely that the beam dilution
will affect the line luminosities and then the derivation of column densities.
Beam dilution can be more easily disentangled from excitation effects when
several lines of different energy levels and critical densities happen at close
by frequencies so that the beam dilution is identical for these lines. {However,
we cannot derive source properties below the lowest measured angular resolution
without an \emph{a priori} analytical model of the source spatial distribution.
The derived column densities are thus beam-averaged to the lowest observed
angular resolution \ie{} $16.8''$ for \metcy{} and $22.7''$ for \cyacet{}. A beam
dilution factor is applied to transistions corresponding to a larger beam. The
derived column densities will then be lower limits, because lower beam filling
factors translate to higher column densities.

\subsubsection{The Bayesian framework}
\TabBayesianPriors{}

In the case of subthermal excitation and optically thin lines, the main
parameters controlling the line intensities (gas density, column density,
temperature, beam dilution) cannot be independently retrieved without additional
information. As stated in Sect.~\ref{sec.intro}, our previous studies of the
Horsehead edge implies a knowledge of the \Ht{} gas density and the gas
temperature for each observed position.

We wish to combine these previously known information about the source with the
observed line shapes in order to determine robust estimates of the column
densities. The formulation of the inverse problem in the Bayesian framework is
the best way to reach our goal. The principles of Bayesian inference can be found
in~\citep{Press.1992,Feigelson.2012}, and a technical description is available in
Appendix~\ref{app.mcmcradex}. In short, the information on the source physical
properties is taken into account by defining informative priors, \ie{}, peaked
distributions of the parameters. In our case, we use standard normal and
lognormal distributions respectively for the temperature and density parameters.
Lognormal laws allow us to span several orders of magnitude for, \eg{}, the
density.

We have no \emph{a priori} information for the column densities and the beam
dilution. In the Bayesian framework, we used ``uninformative'' distributions
(named Jeffreys prior), \ie{} uniform distributions in a given plausible
parameter range. RADEX has hard coded limits of 5 and 25 for $\log_{10} N$. It is
safe to use a smaller interval for the considered species in the Horsehead case.
The chosen statistical law for the column density was thus taken as a uniform
distribution of the logarithm of the column density with 9 and 17 as boundaries.
For the beam dilution, we chose a uniform distribution between 0 and 1.
Table~\ref{tab.priors} summarizes the quantitative information characterizing the
used distributions.

\subsection{Derived results}

\subsubsection{\label{sec.opacity}\metcy{} abundance, beam dilution and line opacities}

\TabBeamDilution{}
\TabFitLVG{} 

We used the collisional coefficients from \citet{Green.1986}. These coefficients
were computed for collisions with He and scaled to \Ht{}. They are computed for
251 levels (31375 coefficients) with upper level energies up to 1150K. We
restricted our computation to the first 185 levels (16650 coefficients) up to
upper level energies of 580 K in order to limit the computation time. We then
checked that using the full set of coefficients for the best fit parameters
yields the same integrated intensities.

The main remaining uncertainty is the quality of the used collisional rate as the
potential energy surface and the dynamical method employed by Green (1986) were
approximate. The rate coefficients of \citet{Green.1986}, computed for Helium in
the Infinite Sudden Order (IOS) approximation, and then scaled by a factor
$\sqrt{{\mu_{\metcy{}-\Ht}}/{\mu_{\metcy{}-\chem{He}}}}=1.38$ to obtain the value
for \Ht{} are expected to be lower than the actual \Ht{} rate coefficients by up
to an order of magnitude. We thus checked the sensitivity of these results to the
variation of the collisional rate coefficients by multiplying them by one order
of magnitude. The absolute abundances varies by less than a factor 3 and the
abundance ratio between the PDR and the dense core positions varies by at most
50\%. Both results are thus robust.

As the hyperfine structure of some \metcy{} lines were resolved, there are two
independent ways to measure the line opacities. The first one explicitly fits the
hyperfine structure using the \texttt{GILDAS} \texttt{HFS} fitting
method\footnote{The \texttt{HFS} fitting method is described in the documentation
of the \texttt{GILDAS/CLASS} software at
\url{http://www.iram.fr/IRAMFR/GILDAS}.}. In our observations, the $(J=5-4,K=2)$
set of hyperfine lines at the PDR position features the best compromise between
signal-to-noise ratio and separation of the hyperfine components to enable a
meaningful fit. The derived opacity is $\tau=1.3\pm1.1$. The low opacity of
\metcy{} is consistent with the non-detection of \chem{^{13}CH_{3}CN} and
\chem{CH_{3}^{13}CN}. The second method is based on opacities modeled by RADEX.
For the same line, the derived opacity is $0.22\pm0.05$. None of the lines
modeled by RADEX have opacities higher than 0.6 (see
Fig.~\ref{ch3cn_pdr_result}). Given the large uncertainties of the HFS
method in the estimate of the line opacities, this value is not taken into
account in the analysis. Nevertheless, from the radiative transfer analysis, we
derive that the \metcy\ lines are optically thin.

The beam filling factors for \metcy{} are summarized in Table~\ref{tab.beam}. At
the PDR position, the beam dilution factors are compatible with a source
structure that is small ($<10\arcsec$) in only one dimension. In the dense core,
only the two lower frequency lines are detected and the beam dilutions have
similar values, larger than 0.5, implying the emission is more extended than at
the PDR position.

The best fit results, which take the excitation by collision with electrons into
account, are summarized in Table~\ref{tab.fit.lvg}. The associated \metcy{}
column densities for the PDR and core positions, are
$N_{\metcy{}}=\sciexp{(6-15)}{12}\pscm$ and
$N_{\metcy{}}=\sciexp{(3-8)}{11}\pscm$, respectively {these values correspond to
the smallest observed beam for each molecule}. This translates into abundances of
\sciexp{(2-4)}{-10} for the HCO peak and \sciexp{(5-12)}{-12} for the \DCOp{}
peak. \emph{\metcy{} is therefore 30 times more abundant in the PDR than in the
dense core.}

\subsubsection{\isometcy{} abundances and \isometcy{}/\metcy{} isomeric ratio}

Only 3 \isometcy{} lines are detected at the HCO peak and none at the \DCOp{}
peak. Including the beam dilution factors, more than 3 unknowns must be
constrained forbidding a complete modeling of the physical parameters of the
source without additional \emph{a priori} assumptions. In addition to the same
prior on the gas density and temperature as for \metcy{}, we assumed that the two
molecules are cospatial, which translates into identical beam dilutions. In
addition, we do not know about any computation of the \isometcy{} collisional
coefficients. We thus used the \metcy{} ones.

With these hypotheses, we derived a column density of $N_{\isometcy} =
\sciexp{(6-25)}{11}\pscm$ at the PDR position and $N_\isometcy\le
\sciexp{5}{11}\pscm{}$ as the 2 sigma upper limit in the dense core.

The \isometcy{}/\metcy{} abundance ratio is $0.15\pm0.02$ and $<0.15$ at the PDR
and the dense core position, respectively. This value is similar to the ratio of
the integrated intensities. This points towards optically thin lines.

\subsubsection{\cyacet{} and \isocyacet{} abundances and isomeric ratio}

We used the collisional coefficients computed by \citet{Wernli.2007} for
para-\Ht{} and Faure et al. (in preparation) for ortho-\Ht{}. The derived
critical densities for collisions of \cyacet{} with \Ht{} are $4\dix{5}-1\dix{6}
\pccm$. In this modeling, we used the standard RADEX prescription, \ie{} the
determination of the ortho-\Ht{}/para-\Ht{} from the gas kinetic temperature
(0.01 and 0.52 for $T_\emr{kin}=25$ and 60\K{}, respectively).

However, we checked that using a fixed ortho-\Ht{}/para-\Ht{} ratio of 3 does not
influence the results above the derived uncertainties.

For \cyacet{}, all 4 lines were detected at 3mm at both positions, while neither
the 2 lines at 2mm nor the 8 lines at 1mm were detected. The derived column
densities and abundances of \cyacet{} are similar at the PDR and dense core
positions with values of \sciexp{(1-5)}{11}\pscm{} and \sciexp{(5-12)}{-12},
respectively.

For \isocyacet{}, none of the millimetric lines were detected in either of the
observed positions. Assuming the same priors on the gas temperature and density
and cospatial emission of \isocyacet{} and \cyacet{}, the derived 2 sigma upper
limits of the column densities of \isocyacet{} are $N_{\isocyacet{}} \le
\sciexp{2}{10}\pscm$ at both position positions.

Since the opacities of all observed lines are low for \cyacet{}, the abundance
ratio is similar to the observed intensity ratio. The 2 sigma upper limit for the
\isocyacet{}/\cyacet{} abundance ratio thus is $0.1$.

\subsubsection{\cycethy{} abundance}

{   
\TabCCCN{}}

Three doublets of \cycethy{} are marginally detected at the PDR position in the
3mm band. As the collision rates for \cycethy{} are not available, we used the
\texttt{CLASS/WEEDS}~\citep{Maret.2011} LTE engine to model these lines. Based on
previous experience~\citep{Guzman.2012,Pety.2012}, we used an excitation
temperature of 10\K{}, and a source structure corresponding to a $6''\times50''$
filament centered on the PDR position. This yields a column density of
\sciexp{(2\pm1)}{12}\pccm{} at the PDR position, \ie{}, an abundance of
\sciexp{(5\pm2.5)}{-11}. No individual line of \cycethy{} is detected at the
dense core position, even though stacking yields a $5\sigma$ detection. We thus
report an $2\sigma$ upper limit on the column density at the dense core position
of \sciexp{1}{12}\pccm{}, \ie{}, a $2\sigma$ upper limit on the abundance of
\sciexp{1.5}{-11}.

\subsection{Comments}

\subsubsection{\Ht{} vs electron excitation}

Table~\ref{tab.fit.lvg} compare the best fit parameters obtained either without
electron excitation or with a fixed electron fraction of $10^{-4}$ as derived for
the PDR position by \citet{Goicoechea.2009a}. Including electron excitation
reduces the derived column density by 40\% for \metcy{}, but by less than the
uncertainty on the column density (\ie{}, $\le25\%$) for \cyacet{}. This
difference between the two molecules is due to a higher critical densities of
\metcy{} lines ($\sim10^6\pccm$) compared to those of \cyacet{} ($\sim10^5\pccm$).

\subsubsection{Shortcomings of rotational diagrams in the case of \metcy{}}

\metcy{} is a relatively complex species. It is often assumed to exist only in
dense environments where it is thermalized. In such environments, rotational
diagrams~\citep{Goldsmith.1999} are then the tool of choice to derive the column
density of \metcy{}. However, we find that \metcy{} is brighter and much more
abundant in UV-illuminated relatively low density gas than in the UV-shielded
dense core. In appendix~\ref{app.rot}, we discuss the use of rotational diagrams
in the case of moderately subthermal excitation such as in the PDR position. The
determination of the column density through the use of a rotational diagram would
have been underestimated by a factor 6 to 33, depending on the observed
\Kq{}-ladder, at the PDR position compared to the escape probability
computations. The density being higher at the dense core position, the rotational
diagram yield the same column density as the escape probability method.

\section{Discussion}
\label{sec.discussion}

Even when taking the excitation by electrons into account, which is significant
in the UV-illuminated gas, \metcy{} is 30 times more abundant in the PDR
(\sciexp{2.5}{-10}) than in the UV-shielded dense core (\sciexp{8}{-12}). In
contrast, \cyacet{} has similar abundance at both positions (\sciexp{8}{-12}),
while \cycethy{}{} is only abundant in the PDR (\sciexp{5}{-11}). In this
section, we discuss the chemistry of \metcy{} and the values of the isomeric
ratios.

\subsection{\metcy{} and \isometcy{} chemistry}
\label{sec:metcy}

\newcommand{\kdiss}{\emm{\kappa_\emr{diss}}}
\newcommand{\kra}{\emm{k_\emr{ra}}}
\newcommand{\kdr}{\emm{k_\emr{e}}}
\newcommand{\kdrtot}[1][]{\emm{k_\emr{e}^\emr{tot#1}}}
\newcommand{\nel}{\emm{n_\emr{e}}}
\newcommand{\paren}[1]{\left(  #1 \right) } \newcommand{\bracket}[1]{\left[  #1 \right] } \newcommand{\Av}{\emm{\emr{A_v}}} 
The major gas phase route to \metcy{} and \isometcy{} is thought to be a
radiative association~\citep{Huntress.1979,Bates.1983,Leung.1984}
\begin{equation}
  \chem{CH_{3}^{+}}+\chem{HCN}/\chem{HNC} 
  \rightleftharpoons \chem{C_{2}H_{4}N^{+\star}} 
  \rightarrow \chem{C_{2}H_{4}N^{+}} + h\nu, 
\end{equation}
with a reaction constant \kra,
followed by the ion-electron dissociative recombination reaction
\begin{equation}
  \chem{C_{2}H_{4}N^{+}} + \chem{e^{-}} \rightarrow \metcy/\isometcy + \chem{H}.
\end{equation}
The main destruction route in presence of UV illumination is
\begin{equation}
  \metcy{} + h\nu \rightarrow \chem{CH_{3}} + \chem{CN}.
\end{equation}
In this simplified view, the abundance of \metcy{} is given by
\begin{equation}
  \frac{d[\metcy]}{dt} = \kdr \, \nel \, [\chem{C_{2}H_{4}N^{+}}] - \chi \, \kdiss \, [\metcy],
\end{equation}
where the rate of the dissociative recombination to \metcy{} \citep[]{Vigren.2008}
\begin{equation}
  \kdr = 1.5 \times 10^{-7}\paren{\frac{T}{300}}^{-0.5} \unit{cm^3\,s^{-1}},
\end{equation}
the photodissociation rate \citep[]{Dishoeck.2006} is
\begin{equation}
  \kdiss = 1.56 \times 10^{-9}\unit{s^{-1}}\,\exp(-\gamma\,\Av) 
  \quad \mbox{with} \quad
  \gamma = 1.95,
\end{equation}
and $\chi$ is the UV illumination at the PDR edge. The abundance of
\chem{C_{2}H_{4}N^{+}} is given by
\begin{equation}
  \frac{d[\chem{C_{2}H_{4}N^{+}}]}{dt} = \kra \, n_\chem{CH_{3}^{+}} \, [\chem{HCN}]
  -\kdrtot \, \nel \, [\chem{C_{2}H_{4}N^{+}}],
\end{equation}
where the value of the radiative association rate is typically
$2\,10^{-8}\pccm\unit{s^{-1}}$ at 50\K{}~\citep{Bates.1983}, and \kdrtot{} is
the total rate of the dissociative recombination to \metcy{}, \isometcy{},
and possibly other products. The steady-state solution is
\begin{equation}
  [\metcy] = \frac{\kdr}{\kdrtot}\frac{\kra \, n_\chem{CH_{3}^{+}} \, [\chem{HCN}]}{\chi \, \kdiss}.
\end{equation}
The UV illumination for the Horsehead PDR is $\chi \sim
60$~\citep[][]{Abergel.2003,Draine.1978}. Using the known density profile,
we derive a visual extinction of $\Av \sim2$~mag~\citep[see Fig.~4][]{Guzman.2012}.
At the PDR position, the (pure gas phase) Meudon PDR code indicates that
$[\chem{HCN}] \sim10^{-9}$, and $n_\chem{CH_{3}^{+}} \sim 10^{-5}\pccm$.
With these normalizing factors, we deduce
\begin{equation}
  [\metcy] = 10^{-13}\,
             \bracket{\frac{\kdr}{\kdrtot} \,
                    \frac{\kra}{2\,10^{-8}} \,
                    \frac{n_\chem{CH_{3}^{+}} \, [\chem{HCN}]}{10^{-14}} \,
                    \frac{1.21}{\chi \, \exp(-\gamma\,\Av)}}.
\end{equation}

So the predicted abundance of \metcy{} is $\sim 10^{-13}$ at the PDR position.
This is an upper limit as $\kdr \le \kdrtot$. As the measured abundance of
\metcy{} is $\sim2\,400$ times larger than this predicted abundance, this route
does not seem efficient enough to produce the observed amount of \metcy{} at the
PDR position.

A potential alternative to pure gas phase chemistry is the formation of \metcy{}
on ices through ice photo-processing followed by photo-desorption in the
UV-illuminated part of the Horsehead edge. \citet{Danger.2011} have studied the
formation of \metcy{} by the UV photolysis of ethylamine
(\chem{CH_{3}CH_{3}NH_{2}}) ices. They determined that methyl cyanide could be
formed at 20K with a yield of 4\%. Photo-desorption of small molecules like
\chem{H_{2}CO} \citep{Guzman.2011,Noble.2012} is efficient even at low grain
temperatures (20K) where thermal desorption is absent. Larger molecules created
on the ices have higher photo-desorption thresholds and would stay on the grains.

{\metcy{} has also been observed in shocks \citep[\eg{}
L1157-B1][]{Arce.2008,Codella.2009}. In this case, sputtering of grains by
impacting gas evaporates the \metcy{} formed in ices. However, high abundances of
\metcy{} is not a specific tracer of shock, as shown by the example of the
Horsehead PDR.} 

\subsection{Isomeric ratios}

\subsubsection{\isometcy{}/\metcy{}}

\citet{Defrees.1985} argued that the \isometcy/\metcy{} isomeric ratio depends on
the isomerisation equilibrium which happens when the \chem{C_{2}H_{4}N^{+\star}}
unstable molecule or ``complex'' radiatively loses its energy, provided that the
dissociative recombination does not modify the molecule backbone. From
consideration on the energies of the different isomers of \chem{C_{2}H_{4}N^{+}},
they deduced an \isometcy/\metcy{} isomeric ratio in the $0.1-0.4$ range, \ie{}
very different from the typical HCN/HNC isomeric ratio value of 1 found in the
interstellar medium~\citep{Sarrasin.2010,Mendes.2012}.

From their tentative detection of the \J{4}{3}, \J{5}{4}, and \J{7}{6} lines of
\isometcy{} in Sgr B2, \citet{Cernicharo.1988} found a \isometcy{}/\metcy{}
abundance ratio of $\simeq0.05$. Also in Sgr B2, \citet{Remijan.2005} deduced an
even smaller \isometcy{}/\metcy{} abundance ratio of $0.02\pm0.02$ from their
detection of the \metcy{} and \isometcy{} \J{1}{0} lines. The value we derive for
this abundance ratio in the Horsehead PDR is $0.15\pm0.02$ , \ie{} 2--5 times
higher than these previous results. Our value falls directly in the range
(0.1--0.4) predicted by \citet{Defrees.1985}.

\subsubsection{\isocyacet{}/\cyacet{}}

\isocyacet{} was first detected in TMC-1 by \citet{Kawaguchi.1992}, they find a
\isocyacet{}/\cyacet{} abundance ratio of 0.02-0.05. In CRL618 (a circumstellar
envelope affected by strong UV fields from the central star) \citet{Pardo.2007a}
derive an abundance ratio of 0.025-0.03. Our derived upper limit for this
abundance ratio is compatible with these results.

\section{Summary}

While \metcy{} is a complex nitrile molecule (more than 6 atoms), its lines are
brighter in the PDR than in the dense core. Its linewidths are also narrower in
the PDR than in the dense core. Due to the lower density at the PDR position, the
lines are subthermally excited, implying that a rotational diagram analysis
underestimates the column density by a factor 6-33 depending on the observed
\Kq{}-ladder. In other words, bright \metcy{} lines do not necessarily imply
high densities ($\nH \ga 10^{6}\pccm$) and LTE.

Large velocity gradient radiative transfer methods implies that \metcy{} is 30
times more abundant in the UV illuminated gas than in the UV shielded dense core.
The overabundance of \metcy{} in the PDR compared to the dense core position is
surprising because the photodissociation of this complex molecule is expected to
be efficient. A simple pure gas phase chemical model underestimates the abundance
of \metcy{} in the PDR by a factor of at least a few thousand. { We propose that
\metcy{} gas phase abundance is enhanced when ice mantles of grains are destroyed
(photo-desorption or thermal-evaporation in PDRs, sputtering in shocks).}

We also report the first clear detection of 4 \isometcy{} lines in the millimeter
bands at the PDR position. The \isometcy{}/\metcy{} isomeric ratio of 0.15 is
compatible with the theoretical range of 0.1-0.4 from \citet{Defrees.1985}.

In sharp contrast to \metcy{} and its isomer, \cyacet{} lines are stronger in the
dense core than in the PDR. The \cyacet{} abundance is similar at both positions.
No lines of its isomer \isocyacet{} were detected in either position. The pure
gas phase chemistry of \cyacet{} is more complex than the \metcy{} one, requiring
a detail chemical modeling to understand these results.

\begin{acknowledgements} 
This work has been funded by the grant
ANR-09-BLAN-0231-01 from the French {\it Agence Nationale de la Recherche} as
part of the SCHISM project (http://schism.ens.fr/). J.R.G. thanks the Spanish
MINECO for funding support through grants AYA2009-07304 and CSD2009-00038. J.R.G.
is supported by a Ram\'o{}n y Cajal research contract from the MINECO. VG
acknowledges support from the Chilean Government through the Becas Chile
scholarship program. 
\end{acknowledgements}

\bibliographystyle{aa} \bibliography{/Users/gratier/Documents/Biblio/biblio}
\appendix{} 
\section{\label{app.obs}Observational tables}

This section gathers the observational fit obtained for all the lines studied in
this paper. The line parameters are taken from the CDMS~\citep{Muller.2005a} for
\metcy{}, \cyacet{} and \cycethy{}, and from the JPL~\citep{Pickett.1998} for
\isometcy{}, \isocyacet{} and \chem{HNC_{3}}. Original spectroscopic data come
from \citet{Muller.2009} for \metcy{}, \citet{Thorwirth.2000} for \cyacet{},
\citet{Gottlieb.1983} for \cycethy{}, \citet{Bauer.1970} for \isometcy{},
\citet{Guarnieri.1992} for \isocyacet{}, and \citet{Hirahara.1993} for
\chem{HNC_{3}}. The accuracy is excellent for \metcy{}, \cyacet{}, \isocyacet{}
and \cycethy{} and moderate for \isometcy{} ($\sim0.09$ MHz or 0.27\kms{} at 100
\GHz{}), and \chem{HNC_{3}} ($\sim0.12$ MHz or 0.42\kms{} at 84 \GHz{}).

\TabIsometcyObs{} \TabMetcyObs{} \TabCyacetObs{} \TabCyethy{} \TabIsocyacetObs{} \TabHNCCCObs{} 
\section{\label{app.mcmcradex}Bayesian radiative transfer modeling}

The inputs of the RADEX code are the kinetic temperature ($T_\emr{K}$), the
volume density of the collisional partner\footnote{We neglect the influence of
the He collisional partner.} ($n_\Ht$) and the column density of the computed
species ($N$). Given a set of energy levels and the radiative and collisional
transitions linking them~\citep[both from the LAMBDA database][]{Schoier.2005},
the LVG code computes for each line, its opacity, excitation temperature and flux
assuming a gaussian profile of a given fixed linewidth. We combined this LVG
model with a Bayesian fitting method to determine the optimal physical parameters
of the source. We assume that the observation uncertainties are centered
Gaussians. The observed data $D$ is thus represented by a set of $N$ integrated
intensity and its associated measurement uncertainty,
$D=\{I_{i},\sigma_{i}\}_{i=1..N}$.

With the hypothesis of independent Gaussian centered noise, the likelihood of
having observing the data $D$ given the model parameters
${\bm{\theta}}=\{\theta_{i}\}$ is given by 

\begin{equation} 
L(D|{\bm{\theta}}) =
\prod_{i=1}^{N}\left[\frac{1}{\sqrt{2\pi\sigma_{i}}}\exp\left(-\frac{[I^\emr{obs}_
 {i}-I^\emr{mod}_{i}(\bm{\theta})]^{2}}{2\sigma_{i}^{2}}\right)\right],
\end{equation} 

where i is an index over the N channels. Taking the logarithm, the
equation becomes 

\begin{equation} 
\ln L(D|{\bm{\theta}}) =
-\frac{1}{2}\sum_{i=1}^{N}\ln(2
\pi\sigma_{i})-\sum_{i=1}^{N}\frac{[I^\emr{obs}_{i}-I^\emr{mod}_{i}(\bm{\theta})]^
 {2}}{2\sigma_{i}^{2}}, 
 \end{equation}

where $I^\emr{mod}$ are derived from the
parameters ${\bm{\theta}}$ through the RADEX model. In the Bayesian framework,
the posterior probability distribution of the parameters $p(\bm{\theta}|D)$ is
obtained through the Bayes rule 

\begin{equation} 
p(\bm{\theta}|D) \propto
L(D|{\bm{\theta}})p(\bm{\theta}), 
\end{equation}

where $p(\bm{\theta})$ is the prior probability distribution of the $\theta$
parameter. Through the use of informative prior distribution of the model
parameters, it is possible to break model degeneracies. \FigPDRDDParameters{} The posterior probability function can have a complicated surface with more than
one maxima, in order to identify the best set of parameters (\ie{}, the posterior
probability function around the global maxima) numerous tools have been
developed. We use a Markov Chain Monte Carlo (MCMC) method for sampling the
posterior probability function, { specifically {\tt emcee}
\citep{Foreman-Mackey.2013}, a MCMC Python implementation using the
affine-invariant ensemble sampler presented in \citet{Goodman.2010}.} This
sampling method enables us to have the posterior probability distribution as an
equilibrium-sampling distribution. With the set of sampling values of the
parameters it is then possible to compute marginalized one dimension probability
distribution functions for each individual parameters. A central tendency (\ie{}
mean, median or histogram maximum) gives the most probable a posteriori parameter
value and a confidence interval can be directly computed from the probability
distribution of the parameters.

As an example, Figs~\ref{ch3cn_pdr_2dparameters} and~\ref{ch3cn_pdr_result} shows
the results of the modeling of the \metcy{} emission at the PDR position.
Figure~\ref{ch3cn_pdr_2dparameters} shows 2d posterior distributions of the model
parameters. The marginalized 1d probability distribution functions of each
parameter, which are displayed along the diagonal, are integrated over all the
other parameter axes. Figure~\ref{ch3cn_pdr_result} gathers the 1d probability
distribution functions of the RADEX results: 1) the integrated intensity, and 2)
the line opacity as a function of the \K{} number for different (\Jo{J+1}{J})
\Kq{}-ladder.

\FigPDRIntensitiesTau{}

\section{\label{app.rot}Shortcomings of rotational diagrams 
  in the case of \metcy{}}
\FigRotDiagPDR{} \FigRotDiagCore{} 
In this section, we quantify the error made when using rotational diagrams to
derive column densities and abundances in moderately sub thermal excitation
regimes such as those found in the PDR position ($\sim \sciexp{6}{4}\pccm{}$). We
built the rotational diagrams assuming that the line emission is optically thin,
\ie{} we did not correct the measured column densities of the upper levels for
opacity. Figs.~\ref{ch3cn_rotdiag_pdr} and \ref{ch3cn_rotdiag_core} shows the
results. Two points stand out. First, the rotational temperatures derived from a
fit of all the lines are lower than the kinetic temperature. However, the derived
temperatures gets closer to the kinetic temperature when fitting the transitions
with different \Jq{} levels separately (one fit per panel in
Figs.~\ref{ch3cn_rotdiag_pdr} and \ref{ch3cn_rotdiag_core}). This comes from the
fact that different \Kq{}-levels at constant \Jq{} are not radiatively coupled
(cf.\ Eq.~\ref{eq:transitions}).

{When fitting all the lines simultaneously, the rotational diagram derived
column densities are underestimated by a factor 20 compared to the LVG derived
ones for the PDR and a factor 2.5 for the dense core. When the \Kq{}-ladders are
fitted independently, the column densities increase when the associated \Jq{}
level decreases.} Indeed, the high \Jq{} levels are more difficult to thermalize
because their critical densities are higher. As a consequence, the derived column
densities is more and more underestimated as \Jq{} increases. In the Horsehead
PDR case, the column density derived from the rotational diagram for the lowest
\Jq{} level (5-4) is still six times lower than the column density derived by the
escape probability radiative transfer modeling. In contrast, the rotational
diagram and the RADEX methods yields the same column density at the dense core
position. This is expected as the \metcy{} level population will be closer to
thermal equilibrium in the higher density core.

Three different situations happen when using the rotational diagram method to
determine the column density of \metcy{} and/or the gas kinetic temperature.

\begin{enumerate}
\item When the level populations follow an LTE distribution (\eg{} the gas
  density is higher than a few $10^{5}\pccm$ for the J=5-4 \Kq{}-ladder)
  both the gas kinetic temperature and the \metcy{} column density can be
  derived accurately. This approximately corresponds to the case of the
  dense core position in our study.
\item When the excitation is slightly subthermal (\eg{} the gas density is
  in the range $(4-8)\dix{4}\pccm$ for the J=5-4 \Kq{}-ladder), the gas
  kinetic temperature can be obtained by fitting only the corresponding
  \Kq{}-ladder lines that are close to thermalization. However, the column
  density will be underestimated. This corresponds to the case of the PDR
  position in our study.
\item When the excitation is strongly subthermal (\eg{} the gas density is
  lower than $4\dix{4}\pccm{}$ for the J=5-4 \Kq{}-ladder), both the gas
  kinetic temperature and the \metcy{} are underestimated. It is then
  necessary to study the excitation with more advanced methods like escape
  probability methods.
\end{enumerate}

\section{\label{app.largefig}Methyl Cyanide and Isocyanide
spectra}

\FigMetcySpectraSideway{}
\FigIsometcySpectraSideway{}

\end{document}